\documentclass[traditabstract]{aa} 

\usepackage{graphicx}
\usepackage{rotating}
\usepackage{epsfig}
\usepackage{float} 
\usepackage{subfigure}
\usepackage[para,online,flushleft]{threeparttable}
\usepackage{natbib}
\bibpunct{(}{)}{;}{a}{}{,} 
\usepackage{dcolumn}

\newcolumntype{.}{D{.}{.}{-1}}
\newcolumntype{;}{D{;}{.}{7}}

\usepackage{color}

\begin{document}

\title{Polarized near-infrared light of the Dusty S-cluster Object (DSO/G2) at the Galactic center\thanks{Based on 
NACO observations collected between 2004 and 2012 at the Very Large Telescope (VLT) of the 
European Organization for Astronomical Research in the Southern Hemisphere (ESO), Chile.}}
\subtitle{}

\author{B. Shahzamanian\inst{1,2},  A. Eckart\inst{1,2}, M. Zaja\v{c}ek\inst{1,2}, M. Valencia-S.\inst{1}, N. Sabha\inst{1}, \\L. Moser\inst{1,2}, M. Parsa\inst{1,2}, F. Peissker\inst{1}, C. Straubmeier\inst{1}
  }

   \institute{ I.Physikalisches Institut, Universit\"at zu K\"oln,
              Z\"ulpicher Str.77, 50937 K\"oln, Germany\\
       \email{shahzaman@ph1.uni-koeln.de, eckart@ph1.uni-koeln.de}
         \and
             Max-Planck-Institut f\"ur Radioastronomie, 
             Auf dem H\"ugel 69, 53121 Bonn, Germany           
  }

\date{Received:/ Accepted:  }


\abstract {We investigate an infrared-excess source called G2 or Dusty S-cluster Object (DSO), which moves 
on a highly eccentric orbit around the Galaxy's central black hole, Sgr~A*. 
We use, for the first time, near-infrared polarimetric imaging data to determine the nature 
and properties of the DSO and obtain an improved $K_\mathrm{s}$-band identification of this source 
in median polarimetry images of different observing years. The source started to deviate from 
the stellar confusion in 2008, and it does not show any flux density variability over the years we analyzed it. We measured the polarization degree and angle of the DSO between 2008 and 2012 and conclude, based on the significance analysis on polarization parameters, that it is an intrinsically polarized source ($>20\%$) with a varying polarization angle as it approaches the position
of Sgr~A* . The DSO shows a near-infrared excess of $K_{\rm{s}}-L' > 3$ that remains compact close to the pericenter of its orbit. Its observed parameters and the significant polarization obtained in this work show that the DSO might be a dust-enshrouded young star, forming a bow~shock as it approaches the super massive black hole. The significantly high measured polarization degree indicates that it has a non-spherical geometry, and it can be modeled as a combination of a bow~shock with a bipolar wind of the star. We used a 3D radiative transfer model that can reproduce the observed properties of the source such as the total flux density and the polarization degree. We obtain that the change of the polarization angle can be due to an intrinsic change in the source structure. Accretion disk precession of the young star in the gravitational field of the black hole can lead to the change of the bipolar outflow and therefore the polarization angle variation. It might also be the result of the source interaction with the ambient medium.

}


\authorrunning{B. Sh. et al.} 
\titlerunning{}
\maketitle
\section{Introduction}
\label{section:Introduction}

Since 2012, the focus of Galactic center (GC) observations has been set on investigating an infrared (IR) excess source detected by \cite{Gillessen2012} as a fast-moving object approaching the position of the central supermassive black hole (SMBH) of the Milky Way, Sagittarius A* (Sgr~A*). It has been interpreted as a combination of dust and core-less gas cloud called G2 \citep{Gillessen2012, Gillessen2013a, Pfuhl2015} and also DSO, standing for Dusty S-cluster Object \citep{Eckart2013}. This source moves on a highly eccentric orbit and passed its closest approach to the SMBH in May 2014 \citep{meyer2014,Valencia2015}. Given the short distance of its periapse, it has been suspected that it might produce extraordinary accretion events on to the galaxy's central black hole \citep[e.g.][]{Shcherbakov2014, Abarca2014, Scoville2013, Sadowski2013}.

If the DSO is a pure gas cloud of a few Earth masses \citep{Gillessen2012}, it might have formed in the stellar cluster, possibly within the disk of young stars at a distance of few arcseconds from the GC. After forming there, it might have moved on its current remarkably eccentric orbit by gravitational interaction with massive stars \citep{Murray-Clay2012, Scoville2013}. This scenario must have happened recently (1990-2000), therefore it should have been observed during the total time of its existence. As a consequence, the pure gas scenario seems unlikely, and several authors have proposed scenarios suggesting the presence of a central star for this source \citep[e.g.,][]{Murray-Clay2012, Eckart2013, Scoville2013, Ballone2013, Phifer2013, Zajacek2014, Witzel2014, Valencia2015}. \cite{Scoville2013} proposed that the DSO is a T~Tauri star that was formed in the young stellar ring and then inserted into its current orbit. They suggested that a very dense bow~shock is produced for the T~Tauri star wind and modeled it numerically. \cite{Valencia2015} also discussed that the bright observed Br$\gamma$ emission of the DSO with a large line width might be the result of infalling material shaping a disk around the central star, which may be a T~Tauri star with an age of $\sim10^5$ yr. Considering the stellar nature, DSO would not be disrupted when reaching its closest point to the SMBH and did not need a recent formation. Using hydrodynamical simulations, \cite{Jalali2014} have shown that young stars could form very close to SMBHs within small molecular clumps on eccentric orbits around the black hole. They showed that for such orbital configurations, the gravitational potential of the SMBH and orbital (geometrical) compression increase the density of cold gas clumps to reach the threshold values suitable for star formation (see also \cite{mapelli2016} for a recent review).

The $L$-band observations of DSO/G2 close to the peribothron support the idea of the compactness of this source, which means that the source cannot be a pure gas cloud \citep{Ghez2014, Witzel2014}. \cite{Eckart2013} revealed the first $K_\mathrm{s}$-band identification of \textcolor[rgb]{1,0.501961,0}{\textcolor[rgb]{1,0.501961,0}{\textcolor[rgb]{0,0,0}{the} \textcolor[rgb]{0,0,0}{DSO}} }with a magnitude of $\sim$18.9 from the ESO Very Large Telescope (VLT) continuum imaging data. Using the spectral decomposition of this source, they obtained an upper limit of $\sim 30~L_{\odot}$ for its luminosity. The $H$, $K_\mathrm{s}$, and $L$-band continuum measurements can be matched either by an unusually warm dust component at a temperature of 550-650 K or by a stellar source enclosed in the dust at a temperature of $\sim$ 450 K \citep[see Fig.~15 in][]{Eckart2013}. The $H-K_\mathrm{s} > 2.3$ color limit supports the scenario that the DSO is a dust-embedded star and not a core-less cloud of gas and dust \citep{Eckart2013}.

The mass of this object is higher than what was assumed for a pure gas source, but lower than the typical mass of S-cluster stars ($20 M_{\odot}$). \cite{Valencia2015} reported NIR observations of the DSO during its approach to the SMBH at the GC, which were carried out with SINFONI at the VLT from February to September 2014. They detected spatially compact Br$\gamma$ line emission from the DSO before and
after its peribothron passage and also a Br$\gamma$ line width increase, which may indicate that the DSO is a young accreting star with a dust envelope. The observational data were used to obtain the orbital parameters of this object. Comparable to the previous estimates \citep[e.g.,][]{Meyer2014a}, \cite{Valencia2015} obtained a peribothron distance of about $163\pm16$ AU with a half-axis length of about 33 mpc and an ellipticity $e=0.976$.
When the DSO reaches the peribothron, tidal stretching and disruption of the envelope lead to a velocity dispersion enhancement in the accretion flow toward the central star \citep{Eckart2013, Zajacek2014}. Based on a study by \cite{Witzel2014}, the $L$-band emission of the DSO compared to the Br$\gamma$ emission measured by \cite{Pfuhl2015} is more compact. This shows that the $L$-band emission originates from an optically thick dust envelope around a central star, while the Br$\gamma$ emission is coming from the hot gas that is externally heated by ionized photons of the stars close to the DSO. However, \cite{Valencia2015} did not find evidence for significantly extended and tidally stretched Br$\gamma$ emission. The extra emission of Br$\gamma$ close to the DSO position is not connected to the DSO and is most likely emitted from other sources in the field (Peissker et al., in prep.). The DSO is not the only infrared excess source in the S-cluster of the GC, and there are more dusty sources in this region \citep{Eckart2013,Meyer2014a}. These sources might be dust-enshrouded pre-main-sequence stars that form a bow shock
ahead of their path when they move through the medium with a supersonic speed. Other candidates have also been observed in the radio continuum observations of the GC \citep{yusef2015}. \\

Imaging polarimetry is a powerful technique for studying dusty environments such as core-less dusty objects and/or circumstellar dusty regions. The analysis of polarization allows us to quantitatively
evaluate the object geometries and the dust properties. Intrinsic polarization can be generated only if the system is not symmetric. The asymmetry can occur when the radiation field of the star is not isotropic as a result of a geometric distortion, for instance, when the star develops a bow~shock ahead of its path, or when its photosphere surface brightness is not uniform, or in other
words, is influenced by bright spots. Therefore, supplementary to considering the continuum and line emissions from the DSO, studying the light polarization can be very helpful in determining the nature and properties of this source. If the DSO is a bow~shock, the polarization is determined by the bow~shock morphology. Subsequently, the E-vectors are predicted to be perpendicular to the direction of motion if the medium is homogeneous. If the dust shell surrounding the DSO is a disk, then the resulting polarization depends on the disk inclination.

In this paper we analyze the NIR polarimetric imaging data taken with NACO at the ESO VLT using its Wollaston prism to study the polarization properties of the DSO. In Sect. \ref{section:Observations} we begin with the details about the observations, then describe the data reduction and determine the position of the DSO in the images. In
Sect. \ref{section:results} we present the results of the applied flux density calibration method: light curves, polarimetry measurements, and their statistical analysis. We present the performed radiative transfer model to describe the DSO polarization in Sect. \ref{dso-model}. We discuss the implications of our results in Sect. \ref{section:discussion} and finally summarize the main results of the NIR polarimetry of the DSO in Sect. \ref{section:summary}.

    \begin{figure*}[]
     \begin{center}

        \subfigure{%
            \includegraphics[width=0.45\textwidth]{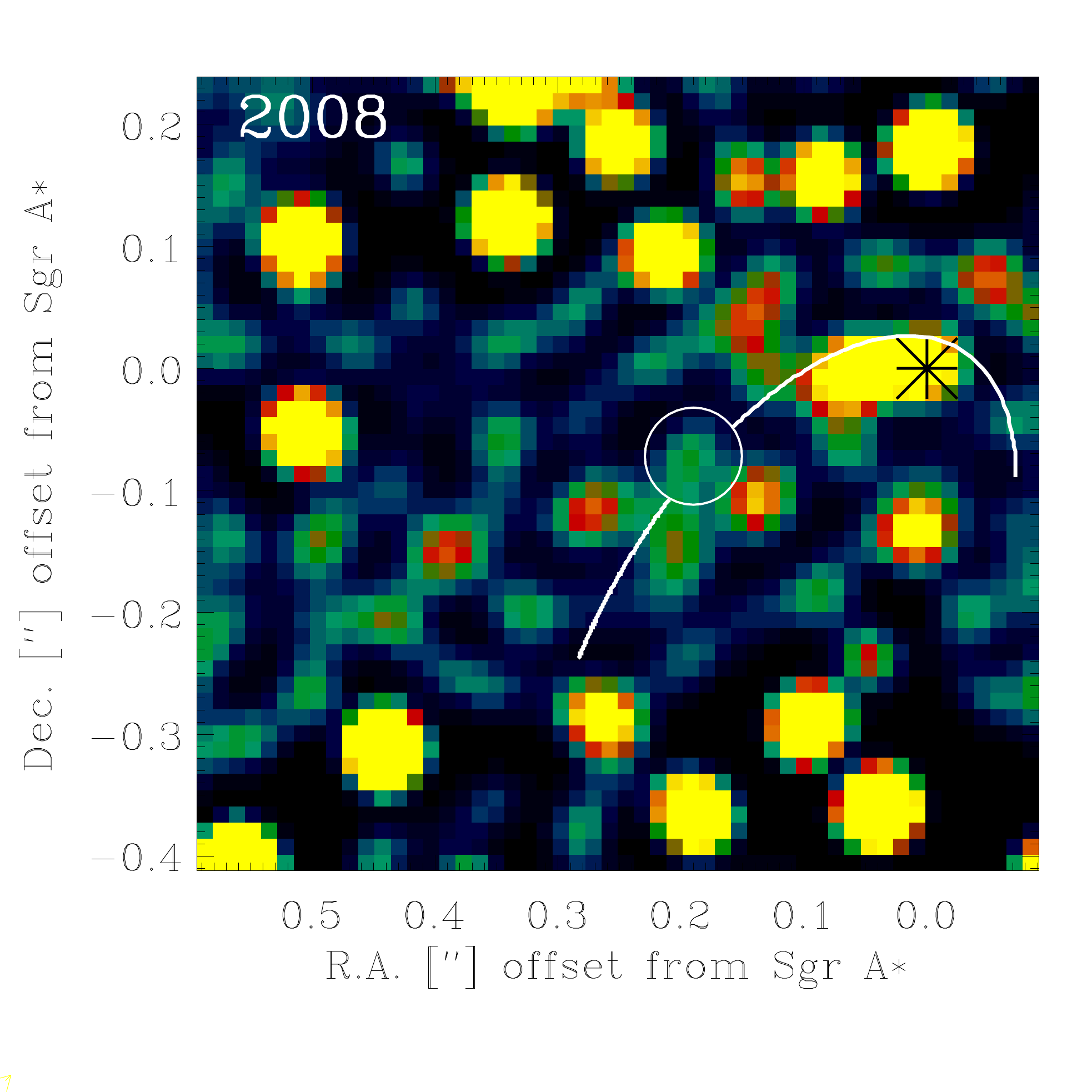}
        }
        \subfigure{%
           \includegraphics[width=0.45\textwidth]{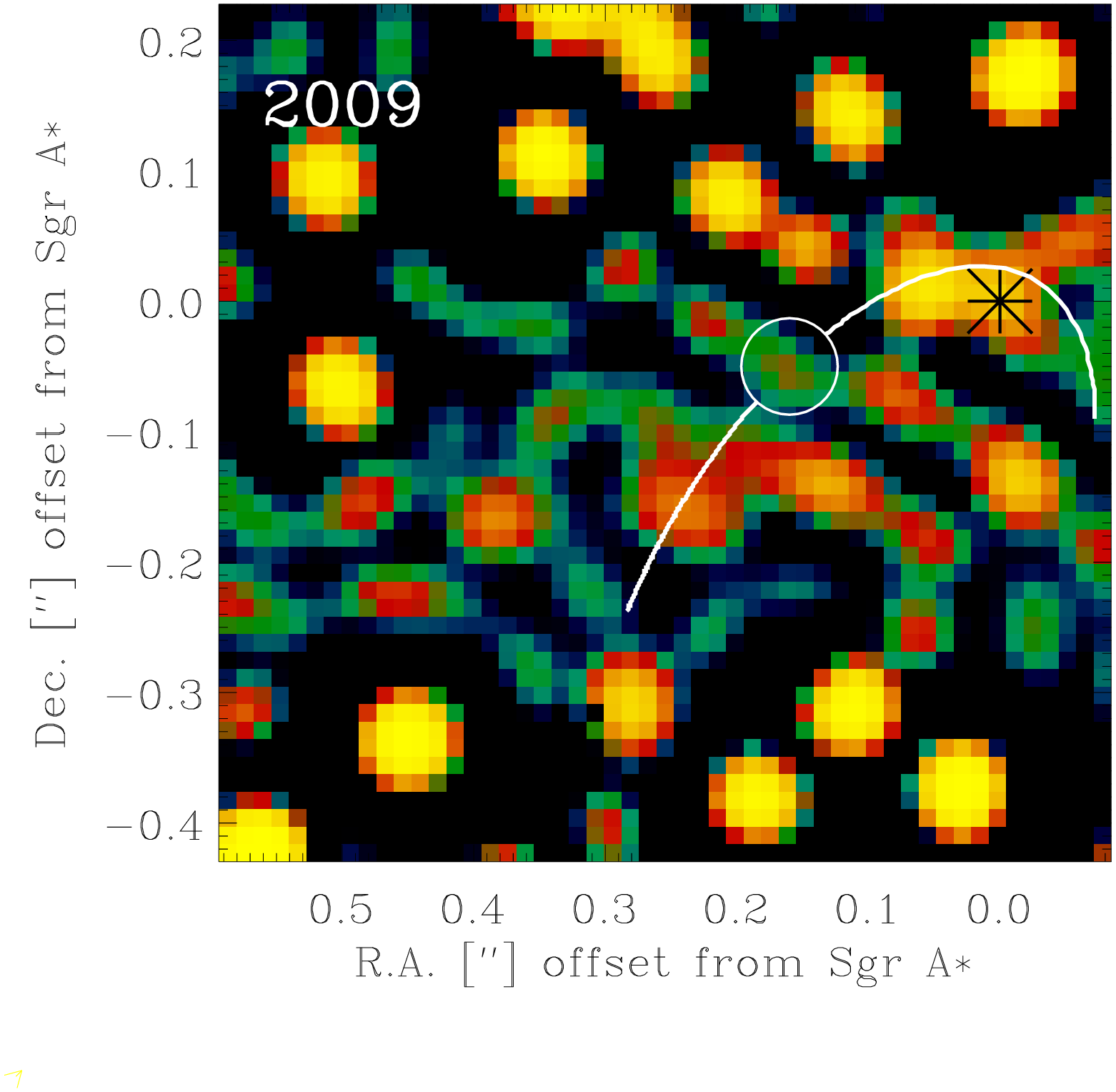}
        }\\
        \subfigure{%
            \includegraphics[width=0.45\textwidth]{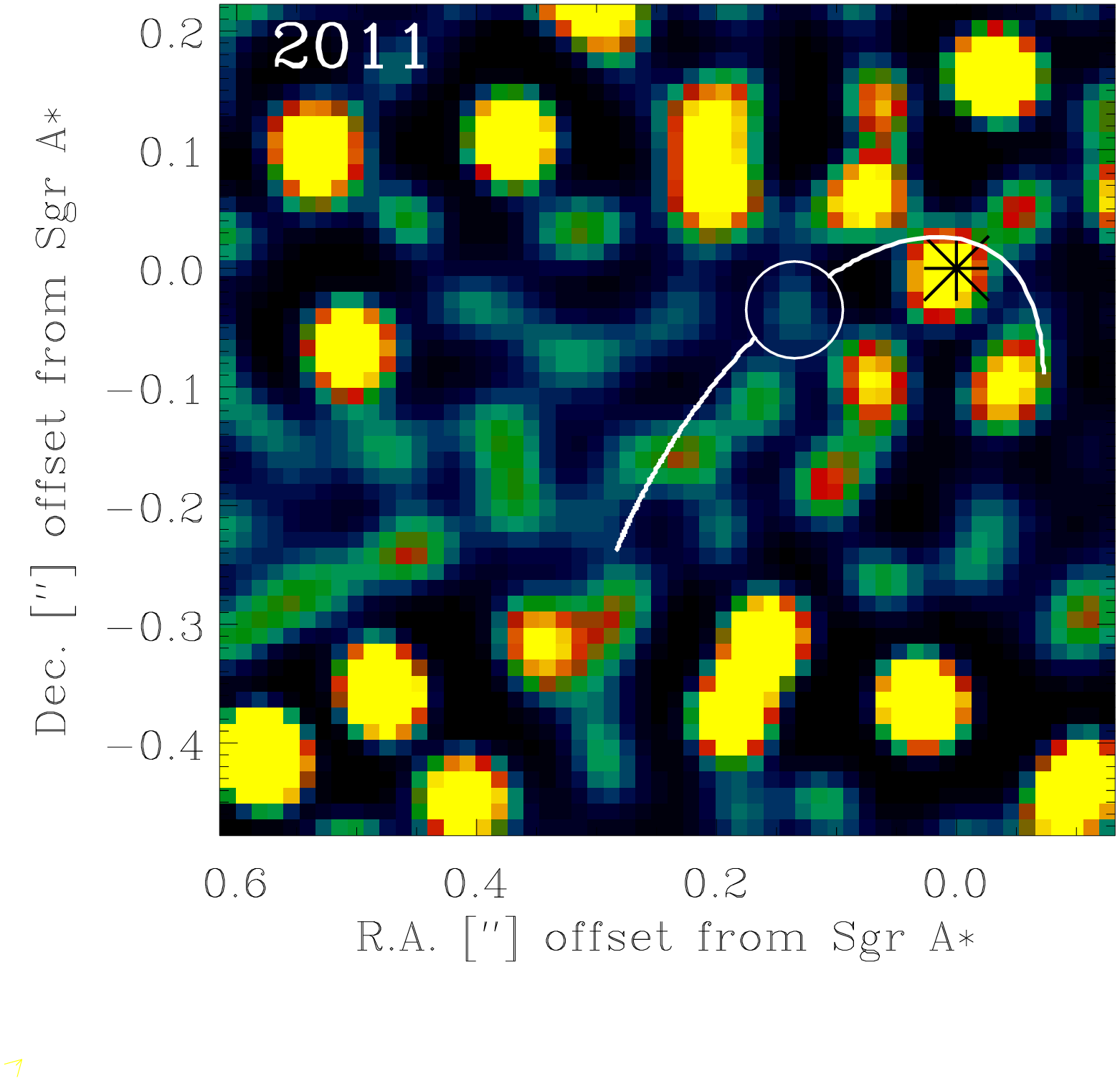}   
        }
        \subfigure{%
            \includegraphics[width=0.45\textwidth]{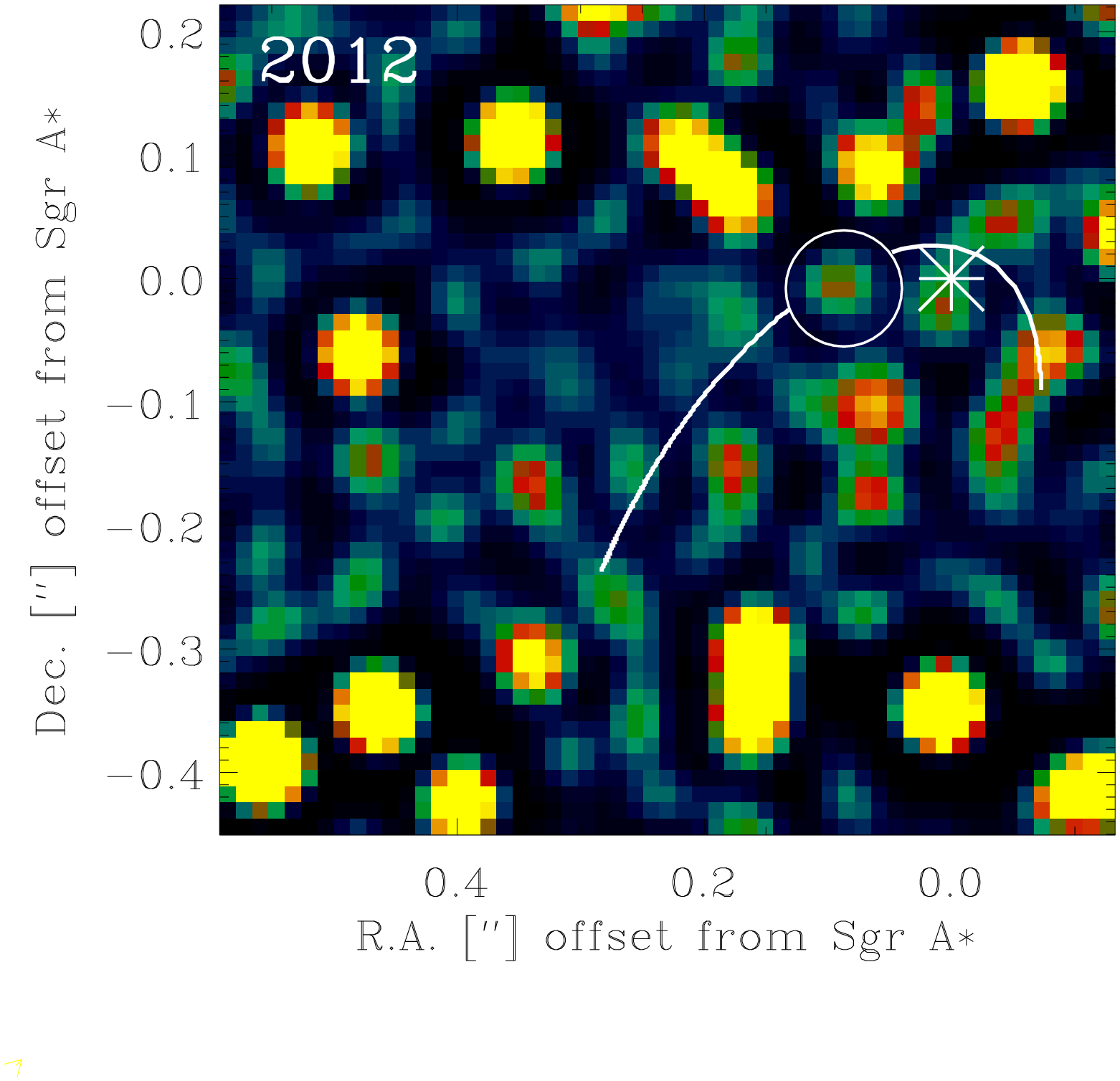}
        }\\

    \end{center}
    \caption[$K_\mathrm{s}$-band deconvolved median images of the central arcsecond of the GC in polarimetry mode]{Final $K_\mathrm{s}$-band deconvolved median images of the central arcsecond at the GC in polarimetry mode ($90^{\circ}$ polarization channel) in 2008, 2009, 2011, and 2012 from top left to bottom right. The position of the DSO is shown by a circle on its orbit.} The asterisk indicates
the position of Sgr~A*. In all the images north is up and east is to the left.
\label{fig:allepochs-dso}
\end{figure*}


\section{Observations and data reduction}
\label{section:Observations}

\subsection{Data set and reduction process}
\label{section:the data set}

The NIR observations have been carried out at the ESO VLT on Paranal, Chile. The data were obtained using the NAOS adaptive optics (AO) module and NIR camera CONICA \citep[together NACO;][]{Lenzen2003, Rousset2003, Brandner2002} at the VLT on Paranal, Chile. The AO loop was locked on IRS7, a supergiant with $K_\mathrm{s}$ $\sim 6.5-7.0$~mag and located $\sim 5.5''$ north of Sgr~A*, using the NIR wavefront sensor. We collected all $K_\mathrm{s}$-band (2.2 $\mu m$) data of the GC taken with S13 camera in 13 mas pixel scale polarimetry mode from 2004 to 2012. The Wollaston prism with the combination of a half-wave retarder plate in NACO provides the possibility of simultaneously measuring two orthogonal directions of the electric field vector. 
We used the reduced data sets as presented in \cite{shahzamanian2015b}. Standard data reduction was applied with flat-fielding, sky subtraction, and bad-pixel correction. All the dithered exposures were aligned using a cross-correlation algorithm \citep[ESO Eclipse Jitter;][]{Devillard1999}. The sky background was measured on a dark cloud located at $713''$ east and $400''$ north of GC. The final sky background was obtained by getting the median of the dithered science epochs. During the observations, the PSF changed because the weather condition varied. Therefore, the quality of each epoch was determined based on the PSF measured from the stars in the field of view at the observing time. We created a data cube from the combination of the best-quality exposures (with seeing $<2''$) of each observing night. To have a final image with longer integration time on the source and higher signal-to-noise-ratio ($S/N$), we obtained the median of the spatial pixels of the combined cube images. In this case, we cannot study the flux density and polarimetry variability on short times of minutes to months. For 2008, we combined all of the observing nights, since the position of the DSO does not change significantly within a few months. Table \ref{table:log} shows observation dates, number of exposures, and integration times of the data sets used for the analysis here before the data were combined.
We used the Lucy-Richardson deconvolution algorithm on
the resulting image of the individual years created from the data cube for
the aperture photometry. The PSF was extracted using the
IDL-based StarFinder routine \citep{Diolaiti2000} from isolated stars close to the DSO position. 

We aligned all the
resulting median cubes of the four polarization channels of the
individual observing years by using a cross-correlation
algorithm. The image was restored by convolving the deconvolved image with a Gaussian beam of a FWHM of about
60 mas.

For the 90-degree channel in 2008, 2009, 2011, and 2012 the
resulting images are shown in Fig.~1. In all the years the DSO
was clearly detected in its continuum emission in all channels
taken with the NACO Wollaston prisms.
To substantiate the detection of the DSO, we also show in the appendix the results of a high-pass-filtered (smooth-subtracted)
image analysis in Fig. A.1 for 2008 and 2012.
In all these images the DSO flux density contribution can clearly be
identified as an individual source component.
While possible PSF contributions have not been cleaned in the high-pass
filtered images, we have conducted our analysis of the DSO emission
using the Lucy deconvolved images.
\begin{table*}
\caption[Galactic center Observations Log]{Galactic center observations log}
\centering
\begin{tabular}{c c c c c}
\hline
\hline 
Date & Start time & Stop time & Number of frames& Integration time\\[0.5ex]
(DD.MM.YYYY)&(UT)&(UT)&&(sec)\\
\hline
\\
25.05.2008 & 06:05:20.32 & 10:35:38.65 & 250 & 40\\
27.05.2008 & 04:52:04.92 & 08:29:38.07 & 184 & 40\\
30.05.2008 & 08:24:33.51 & 09:45:25.69 & 80  &  40\\
01.06.2008 & 06:04:51.56 & 10:10:26.78 & 240 & 40\\
03.06.2008 & 08:37:23.56 & 09:58:58.85 & 80  & 40\\
18.05.2009 & 04:37:55.08 & 10:19:54.10 & 286 & 40\\
27.05.2011 & 04:49:39.82 & 10:27:25.65 & 334 & 45\\
17.05.2012 & 04:49:20.72 & 09:52:57.08 & 256 & 45\\[1ex]

\hline
\end{tabular}
\label{table:log}
\end{table*}

\subsection{Position of the DSO}
\label{dso-position}

We obtained the position of the DSO in each year based on the Br$\gamma$ traced orbit from \cite{Meyer2014a} and \cite{Valencia2015}. Between 2004 and 2007, this source was confused with S63 and could not be distinguished from the background. From 2008 on, it starts to be clear and resolved from stellar confusion. 
The crowded region of the S-cluster may be the reason of this confusion \citep{sabha2012}. An example of the $K_\mathrm{s}$-band identification of the DSO in median polarimetry images for 2008 and 2012 is presented in Fig.~\ref{fig:allepochs-dso}. We used the data set presented in Table \ref{table:log} to investigate the flux density and polarimetry of DSO.

\section{Analysis and results}
\label{section:results}

\subsection{Light curves}
\label{section:lightcurves}


\begin{figure}[b]
     \centering
     \includegraphics[width=\columnwidth]{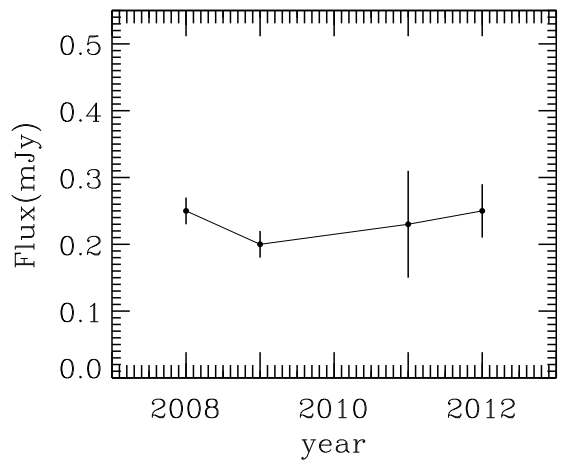}      
    \caption[Light curve of the DSO observed in polarimetry mode]{NIR $K_\mathrm{s}$-band light curve of the DSO observed in polarimetry mode in 2008, 2009, 2011, and 2012.}

   \label{fig:dso_alone}
\end{figure}


We measured the flux density of the DSO in each observing year by aperture photometry using 30~mas radius circular apertures. Flux calibration was made by aperture photometry using circular apertures of 40~mas radius and based on the known flux values of 13 S-stars. Eight background apertures were placed on an area devoid of individual sources to measure the background flux in the deconvolved images. We placed two of these background apertures very close to the DSO position. We used the same stellar calibrators as were presented in Table 1 in \cite{witzel2012} and Fig. 2 in \cite{shahzamanian2015b}. We would like to note that considering the proper background apertures is critical for an accurate flux density estimation of the DSO, since it is a faint source in the $K_\mathrm{s}$ band and background emission can have a prominent effect on its flux density value estimation.
We added the photon counts in each aperture and then added the resulting values of two orthogonal polarimetry channels to obtain the total flux densities. Then we corrected them for the background contribution. This was done for the DSO and the calibrators close to it. The obtained flux densities were corrected for the extinction, that is, the $K_\mathrm{s}$-band extinction correction magnitude of 2.46 derived for the innermost arc-second by \cite{schoedel2010} was adopted.
Subsequently, the light curve was obtained for 2008, 2009, 2011, and 2012, as presented in Fig. \ref{fig:dso_alone}. The uncertainties of the measured DSO flux densities in the light curve were estimated by obtaining the observational noise by setting ten apertures in different positions of the background close to the DSO position. 

The DSO does not show any intrinsic flux density variability in the $K_\mathrm{s}$ band based on our data set. The reason
might be either a period of variability that does not match the time resolution of our data or an irregular variability.  
However, considering the limited data sample, any conclusion on the flux density variability and the effect of stellar contribution on the flux density estimation cannot be made at this point.

\subsection{Polarimetry}
\label{section:polarimetry}

We derived the polarization degree and angle by obtaining normalized Stokes parameters ($I$, $Q$, $U$, $V$) from the observed flux densities,\\

\noindent
\begin{equation}
I = 1\\
\end{equation}
\begin{equation}
Q =\frac{f_{0}-f_{90}}{f_{0}+f_{90}}
\end{equation}
\begin{equation}
U =\frac{f_{45}-f_{135}}{f_{45}+f_{135}}
\end{equation}
\begin{equation}
V = 0,\end{equation}

where $f_{0}$, $f_{90}$ , and $f_{45}$, $f_{135}$ are flux density pairs of orthogonally polarized channels. It is not possible to measure circular polarization in normalized Stokes $V$ since NACO is not provided with a quarter-wave plate.
The circular polarization of stellar sources in the GC is very small \citep{Bailey1984}, therefore we assumed that it can be ignored and set to 0. However, in the case of dusty sources that have a high dust density, circular polarization may be produced by multiple scattering. Polarization
degree $p$ and angle $\phi$ can be obtained as follows:

\noindent
\begin{equation}\label{eq:deg}
p =\sqrt{Q^{2} + U^{2}}
\end{equation}
\begin{equation}
\phi =\frac{1}{2}~\arctan\left(\frac{U}{Q} \right) 
,\end{equation}

\noindent

\begin{table}[b]
\caption[Polarimetry results of the DSO]{Polarimetry measurements of the DSO}
\centering
\begin{tabular}{c c c}
\hline
\hline 
Observing year & $p$ & $\phi$ \\[0.5ex]
\hline
2008 & $30.14\%$ & $ -62.87^{\circ}$\\
2009 & $32.6\%$ & $  42.92^{\circ}$\\
2011 & $29.9\%$ & $  ~18.125^{\circ}$\\
2012 & $37.64\%$ & $ -9.67^{\circ}$ \\[1ex]

\hline
\end{tabular}
\label{table:pol-dso}
\end{table}

where the polarization angle $\phi$ is measured from north to east and samples a range between 0$^\circ$ and 180$^\circ$ or -90$^\circ$ and 90$^\circ$.

\begin{figure}[!t]
  \centering
  \includegraphics[width=\columnwidth]{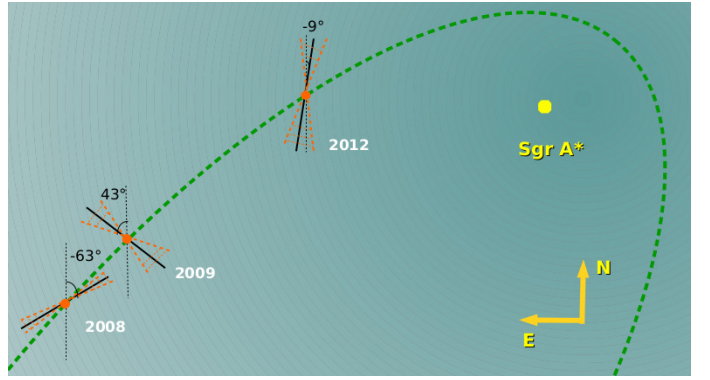}
  \caption[A sketch showing the DSO polarization angle variation]{Sketch of the DSO polarization angle variation when it moves on its eccentric orbit around the position of Sgr~A*  for four different years. The orange shaded areas show the range of possible values of the polarization angle based on our observation and the measurement
uncertainties (see Table 3).}
  \label{fig:sim-4}    
\end{figure}

The instrumental polarization induced in the measured polarization can be corrected for by a model obtained from \cite{Witzel2011} that multiplies a combination of Mueller matrices for different elements of the telescope with the derived stokes vectors. We applied their model to correct for the instrumental polarization effects, where the measured polarization parameters can be estimated with an accuracy of $\sim1\%$ in polarization degree and $\sim5^\circ$ in polarization angle \citep{Witzel2011}.
We show the results of the polarimetry measurements for our data set in Table \ref{table:pol-dso}. Based on the calculated values of $p$ and $\phi$ for different years, the DSO is polarized with a high polarization degree ($p>20\%$) and a polarization angle that changes as the source moves on its orbit. Figure \ref{fig:sim-4} presents a schematic illustration of the DSO polarization angle variation when the source moves on an eccentric orbit around the position of Sgr~A* before it reaches the peribothron.  
Standard error propagation for calculating the polarization uncertainties cannot be applied here since the errors on Q and U are not small compared to the measured values. Moreover, Q and U can be close to zero even if the flux density values in different polarimetry channels are large. In addition, the polarization degree and angle are mainly not Gaussian distributions. \\ 


In general, the effects that are responsible for producing NIR polarization in a GC source can be considered to be intrinsic to the source itself or to foreground effects caused by grain alignment along the line of sight, that is, dichroic extinction. 
Dichroic extinction can also play a significant role in polarization as a local effect for sources that are surrounded by an optically thick dust envelope \citep{Whitney2002}.

\begin{figure*}
     \begin{center}

        \subfigure{%
            \includegraphics[width=0.45\textwidth]{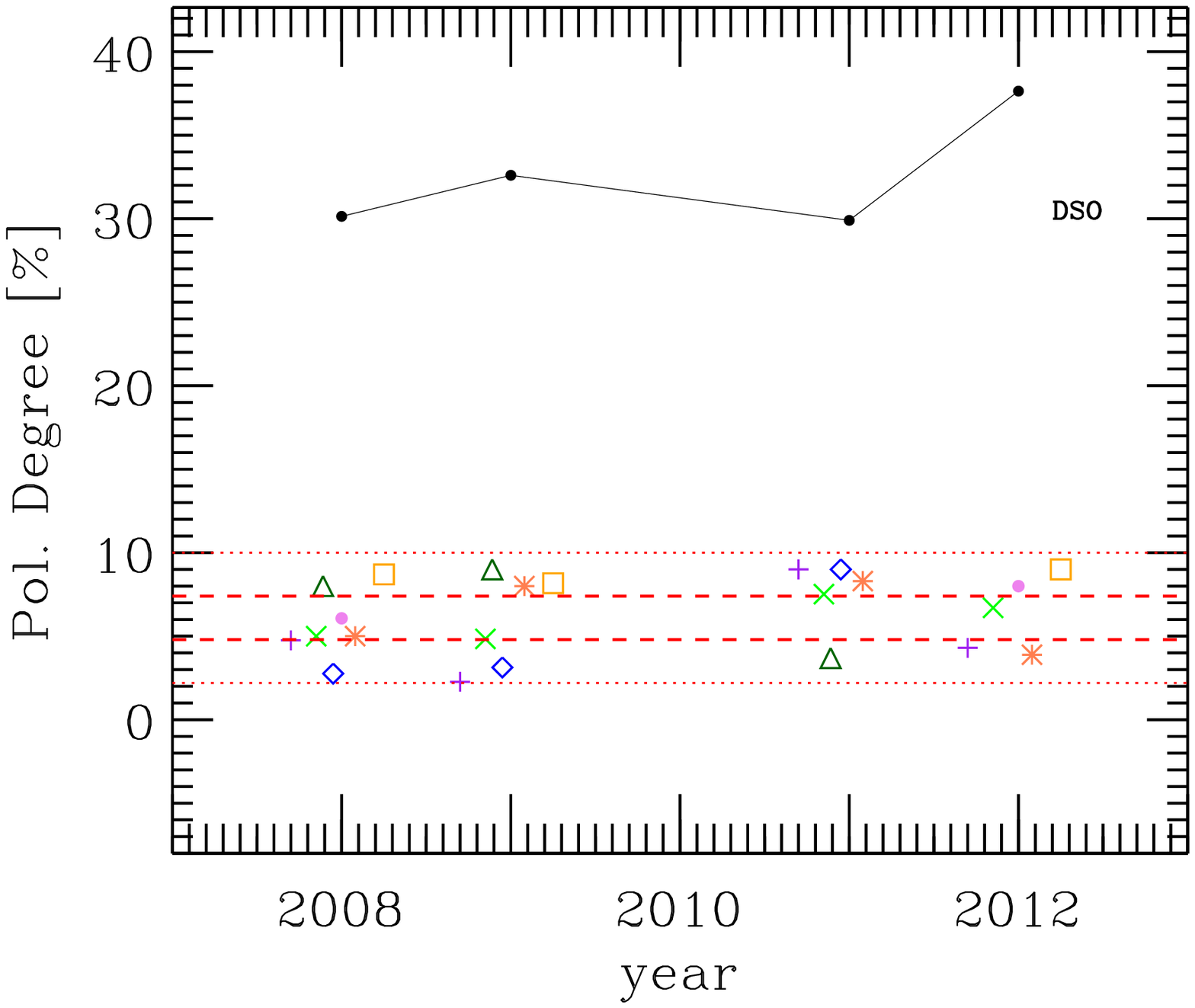}
        }
        \subfigure{%
           \includegraphics[width=0.45\textwidth]{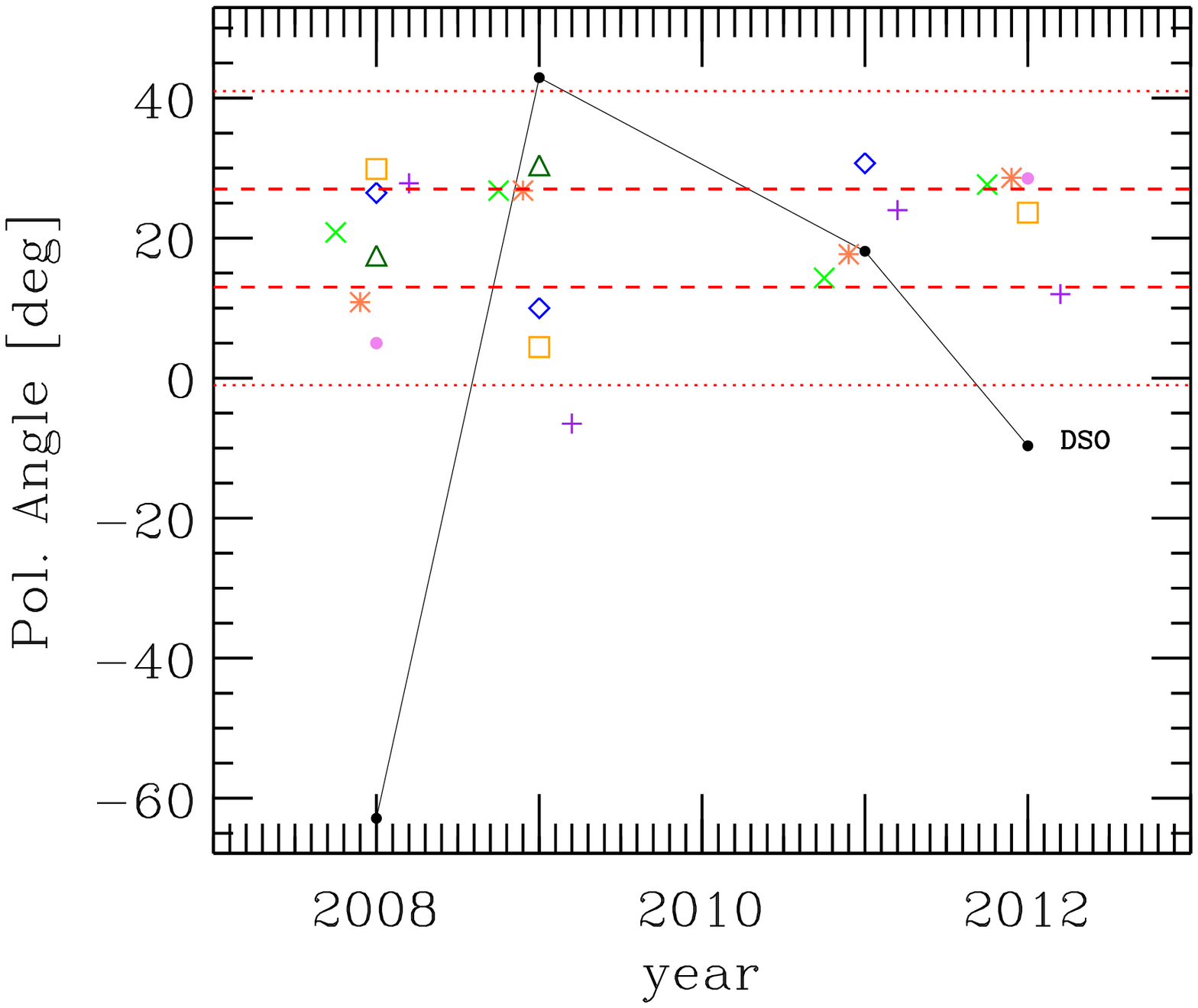}
        }\\

    \end{center}
    \caption[]{Left: Comparison of the polarization degree of the DSO (black dots) with those of GC S-stars located close to the DSO position (S7, S57, S19, S20, S40, S23, and S63; shown with X, asterisk, plus sign, diamond, triangle, violet point, and square, respectively). Right: Comparison of the polarization angle of the DSO (black dots) with those of the S-stars similar to the left panel. In both panels, some of the considered stars are not isolated in some years in which it is difficult to calculate their polarization parameters, therefore we did not show them as data points. The regions between two dashed red lines and dotted lines present the 1 and 3$\sigma$ confidence intervals of the $K_{\rm{s}}$-band polarization degree and angle distributions of the stars reported in \cite{Buchholz2013}, respectively. For the significance and uncertainties of the measured points see Sects.3.3 and 3.4.
      
    }
\label{fig:stars-pol}
\end{figure*}

Figure \ref{fig:stars-pol} presents the polarization degree and angle of the DSO compared to those of several S-stars located close to the DSO position obtained in this work. It also shows the GC foreground polarization from \cite{Buchholz2013}, which is $(6.1 \pm 1.3)\%$ at $20^{\circ}\pm 7^{\circ}$ in $K_\mathrm{s}$
band, widely parallel to the Galactic plane. Based on this figure, the measured polarization degree and angle of the DSO is not the foreground polarization and can be interpreted as intrinsic quantities.\\ 

In addition to the foreground polarization, some sources that are in the Northern Arm and the bow-shock sources have intrinsic polarization in $K_\mathrm{s}$ band as well as at longer wavelengths \citep{rauch2013}. The intrinsic polarization of the stellar sources can be produced by the following processes: emission from aligned dust grains, dichroic extinction, and scattering on spherical grains and/or magnetically aligned dust grains. It is still an open question which of these processes is responsible for producing the observed intrinsic polarization of GC sources. We discuss two of these processes that may produce the DSO polarization
in Sect. 4: scattering and dichroic extinction.\\

\cite{Buchholz2011} found that the $K_{\rm{s}}$-band PSF shows a
polarized structure.
The authors found that for the core of the
PSF (see their Fig~B.2) this effect is basically negligible.
Here, the variations of the intrinsic PSF polarization
amount to a value of only about 1-2\%.
The polarized nature of the PSF is of some importance for extended structures.
We find that the continuum flux distribution of the DSO is unresolved.
However, it is moving within the central stellar cluster.
\cite{Buchholz2011} reported that on the first and second Airy ring
(see their Fig.~B.2) polarization degrees of up to 15\% are possible.
At this location the flux density levels correspond
to less than about 3\% of the peak for a Strehl ratio of
50\%\ in 0.5'' seeing.
In this case the polarized contributions of these neighboring sources
are lower than 2\% of the DSO peak flux.
In 2009 and 2011 the DSO is at even larger distances from similarly
bright sources, hence, the expected polarized flux contributions in these
years is even lower than in 2008.
In 2012 the distance of the DSO to two sources that are about 50 times brighter
is close to the second Airy ring with flux contributions well below 1\%.
Hence, the polarized flux contribution from other sources is well below 15\%.
This implies that in all years the polarization estimates are to more
than 85\% (even more than 98\% in 2008, 2009, and 2011) dominated by the DSO itself.
Therefore, we did not consider PSF polarization variations
in our present investigation.

Assuming the DSO has a bow-shock structure, a comprehensive model of bow~shock polarization using observation results can enable us to analyze the influence of different polarization scenarios (emission, scattering, and dichroic extinction) that play a role in producing the intrinsic polarization of a bow~shock. \\

\subsection{Statistical significance of the measured values}

The polarization degree is a positive defined quantity that becomes biased to
higher values (than the intrinsic $p$ of the source) at low $S/N$ measurements. For this reason, it is possible to measure polarization
degrees that are unphysically higher than 100\% due to the observational uncertainties. Thus,
it is important to keep in mind that the measured (or observed) polarization
degree and angle of any astrophysical object are a good estimation of the
source intrinsic properties only when the $S/N$ is high. 
The DSO $K_\mathrm{s}$-band emission from the yearly stacked images analyzed here have
a medium $S/N$ of $\sim6-10$ in 2008, 2009, and 2012 and a low $S/N \sim2.5$ in
2011. In the first three cases, the distribution of the observed polarization degree
can be described by a Rice function \citep{serkowski1958, vinokur1965}; while in the latter, even this function fails to properly model the high skewness of the polarization degree distribution \citep{simmons1985}.
In turn, the observed polarization-angle distribution depends on the flux-density $S/N$ and on the intrinsic $p$ of the source.


\begin{figure*}[]
  \centering
  \includegraphics[width=1.\textwidth]{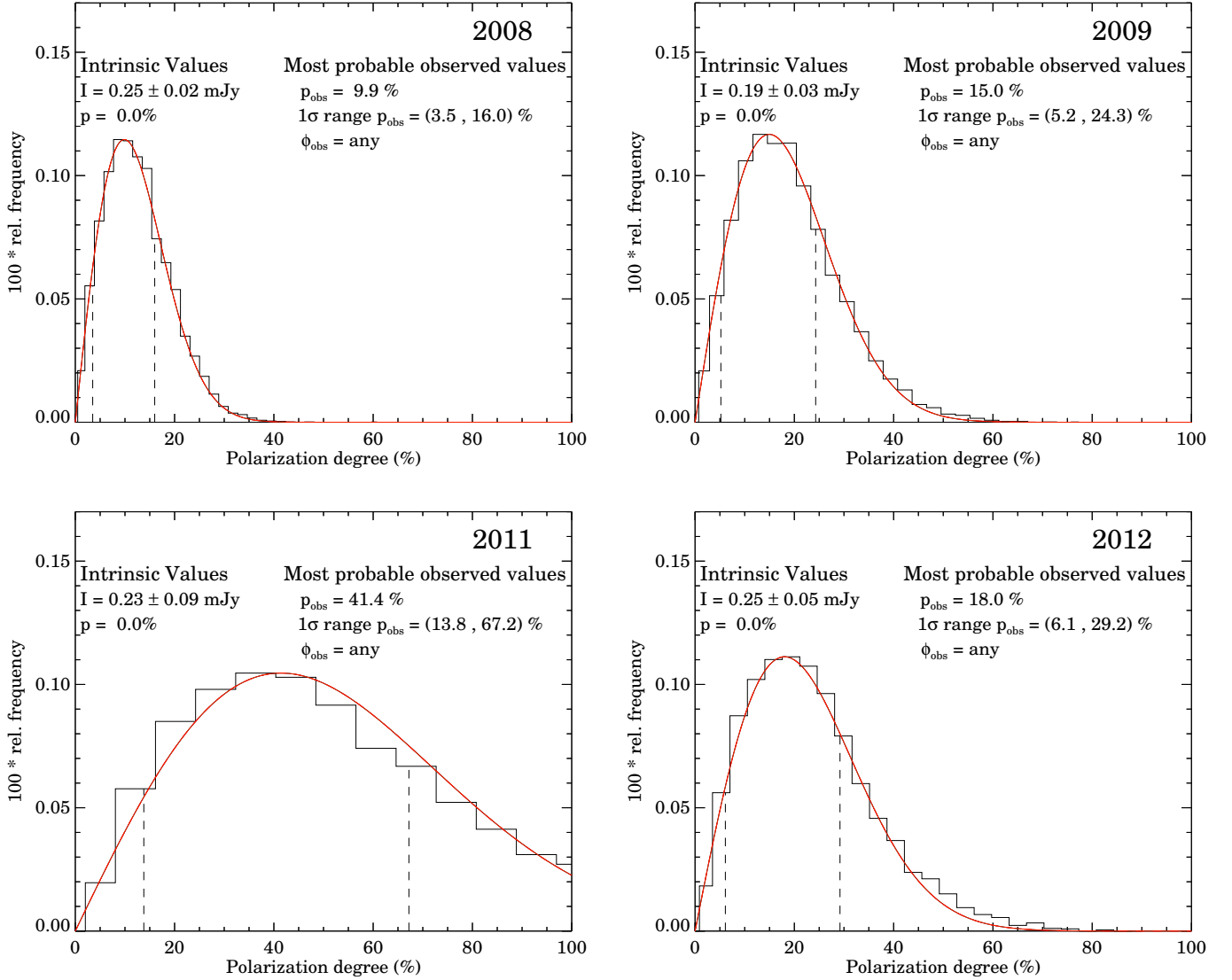}
  \caption{Polarization degree distribution for the observations in 2008, 2009,
2011, and 2012 when considering the null hypothesis ($p=0$\%,
 $\phi=0^{\circ}$), the observed flux density $I,$ and its uncertainty. The
most probable observed polarization degree, $p_{\rm obs}$ in each case
is the mode of the distributions. The range that contains 68\% of the most
probably observed values, $1 \sigma p_{\rm obs}$ is delineated with vertical
dashed lines. The smooth line over the histograms corresponds to the best-fit Rice function.
    }
  \label{fig:figjj}    
\end{figure*}


To estimate the significance of the measured DSO polarization, we assumed as
null hypothesis that the source is intrinsically not polarized, meaning that the
observed polarization angle $\phi_{\rm obs}$ and degree $p_{\rm obs}$ are solely produced by the observational
uncertainties. For this test, we therefore calculated $f_0$, $f_{90}$,
$f_{45}$, and $f_{135}$ from the observed total flux density $I_{obs}$ using
$p=0$\% and $\phi=0^{\circ}$ and assuming the intrinsic flux density to be
the same as the observed $I=I_{obs}$.
We simulated the effect of the noise in the polarization channels by adding a randomly selected amount to each
channel that was drawn from a 
normal distribution with $\sigma(f_{X}) = 0.7\,\sigma(I)$, where $\sigma(I)$
is the uncertainty of the observed total flux density.
Using a Monte Carlo
approach, we calculated 10\,000 times the observed polarization degree and angle
for such a setup using Eqs.\,(1)-(6). The normalized histograms of $p_{\rm obs}$
for the observations in 2008, 2009, 2011, and 2012 are shown in Fig.~\ref{fig:figjj}. 
These distributions represent the probability of measuring certain polarization
degrees under the null hypothesis (i.e., the source is intrinsically not polarized)
given the previously estimated $I$ and their uncertainties.
Figure \ref{fig:figjj} shows that in 2008, where $S/N\approx10$, the most probable observed
polarization degree is $\sim 9$\% with 68/100 of the possible measurements within
 $(4.3, 17.0)$\%.  Under these conditions, measuring a DSO
polarization degree higher than $30$\% when the source is intrinsically not
polarized is only probable in less than 1/100 cases.
In 2009 and 2012, where $S/N\approx6$, the probability of finding an
intrinsically unpolarized source of the same flux density as the DSO, and
$p_{\rm obs} \geq 35$\% is $\sim 10/100$ cases.  
Only in 2011, the low $S/N$ broadens the distribution of $p_{\rm obs}$, 
implying that the intrinsic polarization degree cannot be estimated in this
data set. 
In all tests of the null hypothesis, the observed polarization angle $\phi_{\rm
obs}$ can be any value between $-90^{\circ}$ and $90^{\circ}$ because the
intrinsic polarization degree was set to zero.

We conclude that three of the measurements of the DSO polarization degree are
statistically significant at the confidence levels 0.99 in 2008, and 0.9 in
2009 and 2012. Assuming as starting values for the
simulation $p=6$\% and $\phi=20^{\circ}$, i.e. an intrinsically unpolarized source located at the GC, which, therefore, will display $p$ and $\phi$ values equal to that of the foreground polarization, the significance levels of the polarization measurements change to 0.98 for the 2008, and to 0.88 for 2009 and 2011. This means that intrinsic polarization of the DSO has been detected
in 2008, 2009, and 2012.


\begin{figure*}[]
  \centering
  \includegraphics[width=1.\textwidth]{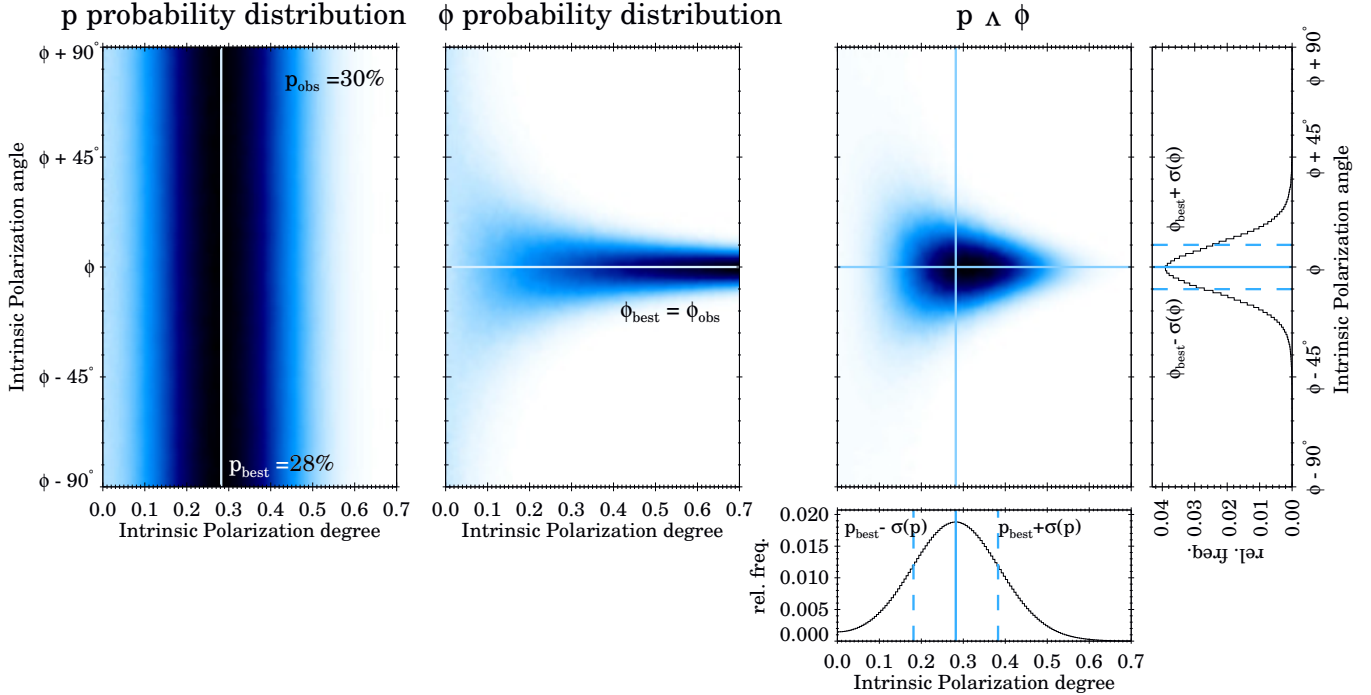}
  \caption{Left: Marginalized distributions of intrinsic polarization degree and angle,
assuming that the observed values are the most probable ones. Here the
case of 2008 data is shown.
Right: Combined probability obtained from the two distributions at the left.
The bottom and right panels show transversal cuts to the combined
distribution at the position of the marked lines.
} 
  \label{fig:figuncer}    
\end{figure*}

\subsection{Measurement uncertainties}

After establishing the statistical significance of the polarization
measurements of three observed epochs, our next goal is to estimate 
the uncertainty of the
intrinsic parameters given the observed values.
Several measurements of an intrinsically non-variable source would be ideal to
assess the uncertainties in the polarization quantities. However, it is unknown
whether the DSO is variable or not, and we can only rely on the stacked images
 of three years. Our alternative approach consists of using
the Monte Carlo simulations presented before to estimate the range of most probable
intrinsic values that would yield the observed DSO polarization. 

We sampled the whole range of possible DSO intrinsic polarization degree and
angle in a grid of $p \in [0.0, 0.7]$ in steps of 0.005, and $\phi \in
[-90^{\circ},90^{\circ}]$ in steps of $2^{\circ}$, calculating each case 
10\,000 times. 
In this way, we constructed two data cubes. One of them has in the first axis
the intrinsic 
polarization degree $p$, in the second axis the intrinsic polarization angle
$\phi$, and in the third the observed polarization degree $p_{\rm obs}$.
The value of $i{\rm-th}$ cell corresponds to the probability of observing $p_{{\rm
obs}, i}$ when the source has intrinsic $p_i$, $\phi_i$, and the observed flux
density (with its uncertainty). 
Similarly, the
second cube has as axes $p$, $\phi,$ and $\phi_{\rm obs}$, and any $i{\rm-th}$
cell equals the probability of observing $\phi_{{\rm obs},i}$ given $p$,
$\phi$ and $I \pm\sigma(I)$. 
From the sheets of $(p, \phi)$ at a constant $p_{\rm obs}$ in one cube, and at a
constant $\phi_{\rm obs}$ in the other, we can deduce the range of intrinsic
polarization  values that most probably yield the observed the polarization
degree and angle.

Figure~\ref{fig:figuncer} shows the $(p, \phi)$ sheets for $p_{\rm obs}=30\%$ and $\phi_{\rm
obs}=-63^{\circ}$, as is the case for the DSO in 2008. As expected, the
probability of measuring certain $p_{\rm obs}$ is independent of the intrinsic
polarization angle of the source and peaks at the most probable intrinsic
$p_{\rm best}$ -- which does not necessarily equal $p_{\rm obs}$.  
In contrast, the probability of observing a particular $\phi_{\rm obs}$
depends on the intrinsic polarization degree and  is symmetric with respect
to $\phi$.  The most probable intrinsic polarization angle is the same as the
observed value $\phi_{\rm best} = \phi_{\rm obs}$.
The 68\% ranges of intrinsic values that most probably produce the observed ones,
$\sigma(p)$ and $\sigma(\phi)$ were
calculated from the figures making transversal cuts at the positions of the most
probable intrinsic $p_{\rm best}$ and $\phi_{\rm best}$, as shown in Fig.~\ref{fig:figuncer} (right).
The results for all years are compiled in Table 3. Under the assumption that the DSO polarization degree is approximately
constant over the studied years, the significance of the polarization degree measurement is
larger than $1/100/,000$. The values for 2011 are reported in the table for completeness, but because of the low $S/N$ of the flux-density measurement that
year, they cannot be interpreted as intrinsic properties of the source.

 \begin{table}
\caption{Most probable DSO intrinsic polarization values.}

\label{tyy}      
\centering     
\begin{threeparttable}                                 
\begin{tabular}{c c c c c c}          
\hline\hline                        
Year & $I \pm \sigma(I) /{\rm mJy}$ & $p_{\rm obs}$ & $\phi_{\rm obs}$ & $p_{\rm best} \pm \sigma(p)$ & $\sigma(\phi)$ \\   
\hline                                   
2008 & $0.25 \pm 0.02$ & $30.1$  &  $117.1$  & $28_{-10}^{+10}$            & $9$            \\ 
2009 & $0.19 \pm 0.03$ & $32.6$  &  $42.9$   & $28_{-10}^{+16}$            & $15$           \\
2011\tnote{a} & $0.23 \pm 0.09$ & $29.9$  &  $18.1$   & $0.5_{-0.5}^{+38}$          & $58$\\
2012 & $0.25 \pm 0.05$ & $37.6$  &  $170.3$  & $32_{-18}^{+18}$            & $15$           \\
\hline                                             
\end{tabular}
     \begin{tablenotes}
          \item[a] These values cannot be interpreted as intrinsic because of
   the low flux-density $S/N$ obtained this year.
          \end{tablenotes}
    \end{threeparttable}
\end{table}

\section{DSO model: a young, supersonic star}
\label{dso-model}

Two mechanisms are responsible for the origin of the polarized flux density: Mie scattering, that is, photon scattering on spherical grains, and dichroic extinction, or selective extinction of photons due to the non-spherical shape of dust grains.
Using the dust continuum radiative transfer modeling, we focused on the process of Mie scattering. The model of the DSO was required
to meet the observed characteristics: total SED constraints--integrated flux densities in H, $K_{\rm{s}}$, L', and M bands, total polarization degree higher than $\sim 20^{\circ}$, and changes of polarization degree and angle along the orbit.\\
 
\begin{figure}[!t]
   \centering
   \includegraphics[height=5cm,width=\columnwidth]{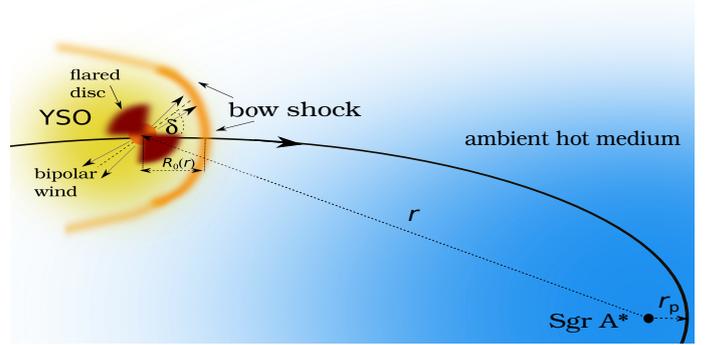}
   \caption{Schematic view of the proplyd-like model of the DSO structure. See the text for a description of the model components.}
   \label{fig:fig_model_DSO}
 \end{figure}

We considered different circumstellar geometries to assess whether they can explain the detected polarized signal and the significant color excess in NIR bands. To test the properties of different constituents, we performed a series of dust continuum radiative transfer simulations using the Monte Carlo code Hyperion \citep{2011A&A...536A..79R}. First, we considered only a stellar source and a bow
shock layer in the calculation, which did not yield a significant polarized
emission. Adding a spherical dusty shell around a stellar
source provided enough extinction to match the NIR excess, but the total polarization degree remained below
$10\%$. Finally, we added a flared disk with bipolar outflows into the density distribution, which caused the source to
deviate more from the spherical geometry (see Fig. \ref{fig:fig_model_DSO}). The polarized
emission in this composite model (star+flared disk+bipolar
outflows+bow shock) is dominated by dust and stellar emission that is scattered on spherical dust grains. The largest
contribution of polarized flux density originates in the region where the outflow intersects the dense bow-shock layer. An overview of different possible circumstellar geometries is provided in Table~\ref{table:geometry}, where the values of the total linear polarization degree in $K_{\rm s}$ band and the color excess $K_{\rm s}-L'$ are included. A high value of the total polarization degree in $K_{\rm s}$ band and an intrinsic NIR excess between $K_{\rm s}$ and $L'$ NIR bands are matched by the composite model star+disk+cavities+dense bow shock, whose properties are studied in more detail.

This scenario resembles proplyd-like objects, which seem to be present in the Sgr~A West region \citep[see][for VLA continuum observations]{2015ApJ...801L..26Y}. Therefore, the DSO might be the first detected and monitored young stellar object in the vicinity of the SMBH that would manifest a very recent low-mass star-formation event close to the GC.  

\begin{table*}[]
\centering
   \resizebox{\textwidth}{!}{  
  \begin{tabular}{c c c c}
  \hline
  \hline
  \\
  \textbf{Geometry} & \textbf{$K_{\rm s}$-band Total Linear Polarization Degree} $\overline{p}_{\rm K,L}$ [\%] & $\mathbf{(K_{\rm s}-L')_{\rm int}}$ (intrinsic) & $\mathbf{(K_{\rm s}-L')_{\rm ext}}$ (with dust extinction)\\
  \\
  \hline
  \\
  Star  & $0$ ($\sim 6\%$ foreground pol. at the GC) & $-0.9$ & $0.4$  \\
  Star+rotationally flattened envelope ($50^{\circ}$ inclination) & $0.2$ & $0.1$ & $1.4$  \\
  Star+flared disk & $3.2$ & $0.3$ & $1.6$\\
  Star+dense bow shock (inclined) & $4.1$ & $1.9$ & $3.2$\\
  Star+spherical dusty envelope+dense bow shock (inclined) & $1.0$ & $1.6$ & $2.9$\\
  Star+flattened envelope+cavities ($90\%$ inclination) & $10.1$ & $1.2$ & $2.4$\\
  Star+flared disk+flattened envelope+cavities ($90\%$ inclination) & $10.5$ & $-1.3$ & $-0.03$\\
  Star+disk+cavities+dense bow shock & $25.0$ & $1.9$ & $3.2$\\
  \\
  \hline 
  \label{table:geometry}
  \end{tabular}
  }
  \caption{Different circumstellar geometries with a list of constituents. Important parameters are the total linear polarization degree in $K_{\rm s}$ band $(2.2\,\rm{\mu m})$ and the color index $K_{\rm s}-L'$. The observed values of the degree and the color index are matched by the composite model star+disk+cavities+dense bow shock that is used in the following analysis.}
\end{table*}

\subsection{Monte Carlo radiative transfer simulations}
 
We performed 3D Monte Carlo radiative transfer calculations using the code Hyperion \citep{2011A&A...536A..79R} to study under which conditions the model of a supersonic, young stellar object would yield the observed DSO properties. The advantage of the Monte Carlo approach is that it is suitable for arbitrary 3D geometry (stellar source surrounded by circumstellar envelope and bow~shock), and we automatically obtain a full Stokes vector $(I,Q,U,V)$, which allows us to easily calculate the total and the polarized flux density. In our simulation runs we set up a spherical polar grid containing $400\times 200 \times 10$ grid points and subsequently added a density grid of gaseous-dusty mixture with a gas-to-dust ratio of $100:1$. A power-law distribution with a slope of $-3.5$ \citep{1994ApJ...422..164K} of spherical dust grains with a radius of between $0.01$ and $10\ \mu\rm{m}$ was considered. The distance to the GC was taken to be $8\,\rm{kpc}$.

The density of the circumstellar envelope was assumed to have a simple radial profile, $\rho_{\rm{disc}}=\rho_0(r/1\,\rm{AU})^{-1}$ with $\rho_{0}=10^{-14}\,\rm{g\,cm^{-3}}$, and the bipolar cavities were assumed to have an opening angle of $90^{\circ}$ and a uniform density of $10^{-20}\,\rm{g\,cm^{-3}}$ \citep{Sanchez2016}. For the bow-shock component and its evolution, we considered the treatment of \citet{Zajacek2016} for the DSO based on the analytical axisymmetric model of \citet{wilkin1996}, keeping the same density and temperature profiles for the hot corona around Sgr~A*. 
A two-dimensional bow-shock shape is obtained by the model of \cite{wilkin1996} 

\noindent
\begin{equation}\label{eq:bowmodel}
R(\theta) = R_{0}\csc\theta\sqrt{3(1 - \theta\cot\theta)},
\end{equation}

\noindent
where $\theta$ is the polar angle from the axis of symmetry, as seen by the star at the coordinate origin. $R_{0}$ is the so-called stand-off distance obtained by balancing the ram pressures of the stellar wind and ambient medium at $\theta = 0,$ and it is given by

\noindent
\begin{equation}\label{eq:offdistance}
R_{0} =\sqrt{\frac{\dot{m_{w}}v_{w}}{\Omega \rho_{a}v_{\star}^{2}}}
.\end{equation}

\noindent
Here, $\dot{m}_{\rm w}$ is the stellar mass-loss rate, $v_{\rm w}$ the terminal velocity of the stellar wind, $\rho_{\rm a}$ the ambient medium mass density where $\rho_{\rm a} = m_{\rm H}n_{\rm H}$ (with $n_{\rm H} \approx n_{\rm e}$), and $m_{\rm H}$ the mean molecular weight of hydrogen, and $v_{\star}$ is the velocity at which the star moves through the medium (i.e., the relative velocity of the star with respect to the ambient medium when the ambient medium is not stationary). The solid angle $\Omega=2\pi(1-\cos\theta_0)$, where $\theta_0$ stands for the half-opening angle of the bipolar outflow, represents the direction into which the stellar wind is blown, outside $\Omega$ there is no outflow \citep{1997ApJ...474..719Z}. For an isotropic stellar wind, $\theta_0=\pi$, the solid angle is naturally $\Omega=4\pi$, whereas for our non-spherical bipolar stellar outflow, $\theta_0=\pi/4$, and the solid angle is then reduced to $\Omega=2\pi(1-\sqrt{2}/2)\approx 1.84$.

The bow~shock consists of two layers: (a) a hot and sparse forward shock, and (b) a cold and denser reverse shock \citep{Scoville2013}. We considered only the colder and denser layer, whose density is about four orders of magnitude higher than for the hot shock \citep{Scoville2013}, and therefore it is more significant in terms of scattering on spherical dust particles. The initial value for the bow-shock density is set to $10^{-16}\,\rm{g\,cm^{-3}}$, in accordance with \citet{Scoville2013} for the distance of $10^{16}\,\rm{cm}$ from Sgr~A*. 
 
For the typical parameters of a young star potentially associated with the DSO, which means a mass-loss rate of $\dot{m}_{\rm{w}}=10^{-8}\,\rm{M_{\odot}\,yr^{-1}}$ and a wind velocity of $v_{\rm{w}}=100\,\rm{km\,s^{-1}}$ , the stand-off distance is of about $10\,\rm{AU}$ \citep{Zajacek2016}, which sets the basic length-scale of the model. The typical angular scale of such an object is $1\,\rm{mas}$ at the distance of the GC, in accordance with the compactness of the DSO observed along the orbit \citep{Valencia2015}. The outer radius of the circumstellar envelope was set to $3\,\rm{AU}$ and the inner radius was set to the typical dust sublimation radius \citep{2002ApJ...579..694M},
 
 \begin{equation}
   R_{\rm{sub}}=1.1 \sqrt{Q_{\rm{R}}}\left(\frac{L_{\star}}{1000\,L_{\odot}}\right)^{1/2} \left(\frac{T_{\rm{sub}}}{1500\,\rm{K}} \right)^{-2}\,\rm{AU}\,,
 \end{equation}
which for the upper limit on the luminosity of the DSO $L_{\star}=30\,L_{\odot}$ and a ratio of dust absorption efficiencies $Q_{\rm{R}}=1$ leads to the estimate of $R_{\rm{sub}}\approx 0.1\,\rm{AU}$. 
 
\subsection{Comparison with observations} 
 
\begin{figure}
  \centering
  \includegraphics[width=\columnwidth]{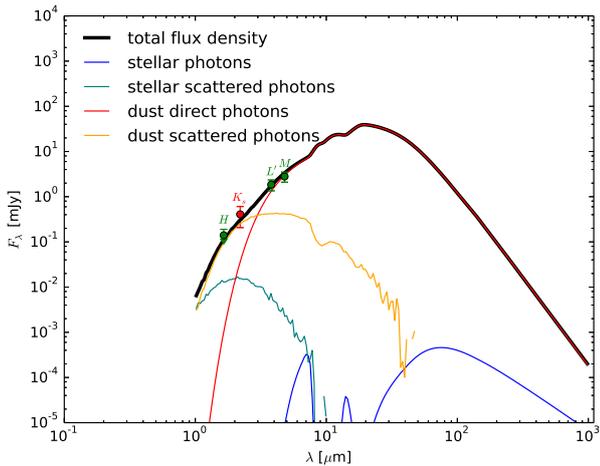}
  \caption{SED for the young stellar object described in the text at the GC. The points label the flux densities inferred from the observations.} 
  \label{fig_sed} 
\end{figure}

The described model of a supersonic, young stellar object with the adopted density distribution can reproduce the inferred flux densities in $H$, $K_{\rm{s}}$, $L'$, and $M$ NIR bands: $F_{H}\leq0.14$~mJy, $F_{K}\sim 0.3$~mJy, $F_{L}\sim 1.9$~mJy, and $F_{M} \sim 2.8$~mJy, respectively \citep{Eckart2013,Witzel2014,Gillessen2012}. In Fig.~\ref{fig_sed} we compare the observed flux densities with the best-fit SED obtained from simulations.  \\

The obtained proper motion of the DSO between 2008 and 2009 is $v_{\alpha} = (-400)\mathrm{km\,s^{-1}}$, $v_{\delta} = (800)\mathrm{km\, s^{-1}}$ and between 2011 and 2012 it is $v_{\alpha} = (-2000)\mathrm{km\, s^{-1}}$, $v_{\delta} = (600)\mathrm{km\, s^{-1}}$. Therefore, $v_{\star}$ changes from $894.42\mathrm{km\, s^{-1}}$ to $2088.06\mathrm{km\, s^{-1}}$. The uncertainty of the proper motion is of about $1\mathrm{mas\, yr^{-1}}$ \citep{eckart2012a}. Using the relative velocity values from 2008 to 2012 in Eq. \ref{eq:offdistance}, the stand-off distance becomes about half of its value \citep[see also Fig. 3 in][]{Zajacek2016}. Moreover, the density profile of the central region based on X-ray data derived in \cite{shcherbakov2010} indicates that the particle number density increases within the
central arcseconds from the SMBH. \\

Owing to the increasing orbital velocity of the DSO and higher ambient pressure toward the pericenter, the bow~shock shrinks and becomes denser. An increment by a factor of four and potentially even more in the bow-shock number density is expected between 2008 and 2012 under the assumption that the bow-shock mass stays approximately constant. This leads to the progressively higher scattering from the non-spherical bow shock for smaller distances from the SMBH.
The increase in the density of the bow-shock layer leads to the progressive increase in polarization degree from an initial $30\%$ in 2008 to almost $40\%$ in 2012, which can be reproduced by radiative transfer simulations of the dust continuum.

The observed change in the polarization angle $\Phi$ is given by the combination of the \textit{\textup{external}} factors (i.e., motion of the DSO through the external accretion flow) and/or the \textit{\textup{internal}} factors (geometry of the circumstellar environment: disk, bipolar wind, and the bow~shock).

\begin{figure}[!t]
  \centering
  \includegraphics[width=\columnwidth]{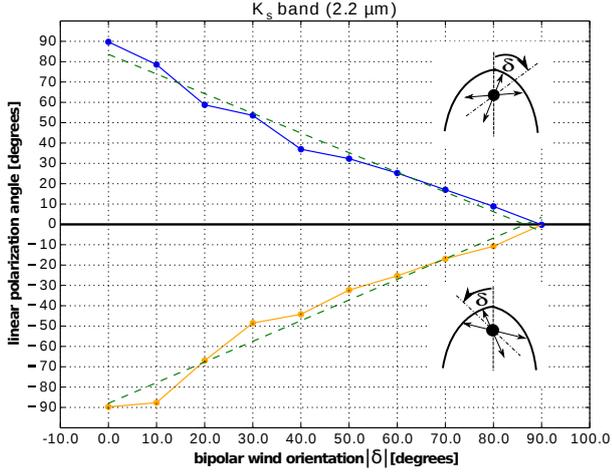}
  \caption{Linear polarization angle as a function of bipolar wind orientation $\delta$. The plot consists of two parts: the upper part shows the dependence for the clockwise rotation of the bipolar wind; the lower part is the same dependence for the counter-clockwise rotation of the bipolar wind. The lines correspond to the linear interpolation of the data; linear regression lines are depicted by green dashed lines.}
  \label{img_delta}
\end{figure}

Using the Monte Carlo radiative transfer calculations, we investigated the effect of the change in orientation of the bipolar wind with respect to the symmetry axis of the bow~shock. The angle between the symmetry axis of the bow~shock and the axis of the bipolar wind is denoted by $\delta$. In our simulations, we gradually increased $\delta$, starting at $0^{\circ}$ (bipolar wind aligned with the bow-shock axis), and we stopped at $90^{\circ}$ (bipolar wind perpendicular to the bow-shock axis). The increment is $10^{\circ}$ , and we performed the simulations in both clockwise and counter-clockwise directions. In Fig.~\ref{img_delta} we plot the dependency $\Phi=f(\delta)$ for the two directions (upper and lower part). In the calculations the bow shock lies in the orbital plane and we observe it from above. The calculated values are represented by points and the lines stand for linear interpolation (blue and orange lines) and the linear regression (green dashed lines).

The basic trend in Fig.~\ref{img_delta} is the following: for the bipolar wind aligned with the bow-shock axis, $\delta=0^{\circ}$, the polarization angle reaches values $\pm 90^{\circ}$; when the angle $|\delta|$ deviates from zero and approaches $90^{\circ}$, the polarization angle approaches zero. Therefore, the change of the orientation of the bipolar wind covers the whole range $(-90^{\circ},+90^{\circ})$. The fitted linear relations are $\Phi_1=-0.97(\pm 0.06)\delta+84(\pm 3)$ (clockwise rotation) and  $\Phi_2=1.02(\pm 0.07)\delta-88(\pm 4)$ (counter-clockwise direction), which in principle may be approximated in the following way: $\Phi \approx -(+) \delta +(-) 90$. The change in orientation of the bipolar wind takes place as a consequence of the torques induced by the supermassive black hole, which leads to its precession
when the circumstellar disk is misaligned with the orbital plane. This naturally affects the inclination of the outflow that originates
in the disk, and the precession takes place on the precession timescale, which is longer than the orbital timescale, $T_{\rm{prec}}>T_{\rm{orb}}$. The wobbling of the disk occurs on timescales shorter than one orbital period, approximately $T_{\rm{wob}}\sim 1/2\,T_{\rm{orb}}$ \citep{2000MNRAS.317..773B}.

In addition, the change in polarization angle may be the combination of the change in internal geometry of the outflow and the external interaction of a supersonic star with an ambient wind with a certain velocity field. When the bipolar wind is aligned with the symmetry axis of the bow~shock, the total polarization angle is perpendicular to the bow~shock symmetry axis. Subsequently, when the supersonic star interacts with the external outflow
or inflow, the bow-shock orientation changes because of the change in relative velocity. This would naturally lead to the corresponding alternation of the polarization angle. \\

\begin{figure}
  \centering
  \includegraphics[width=\columnwidth]{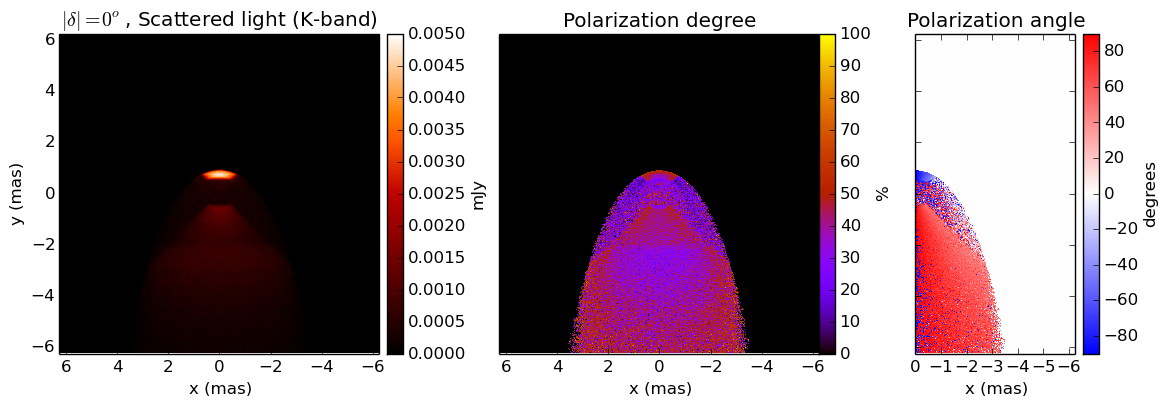}
  \includegraphics[width=\columnwidth]{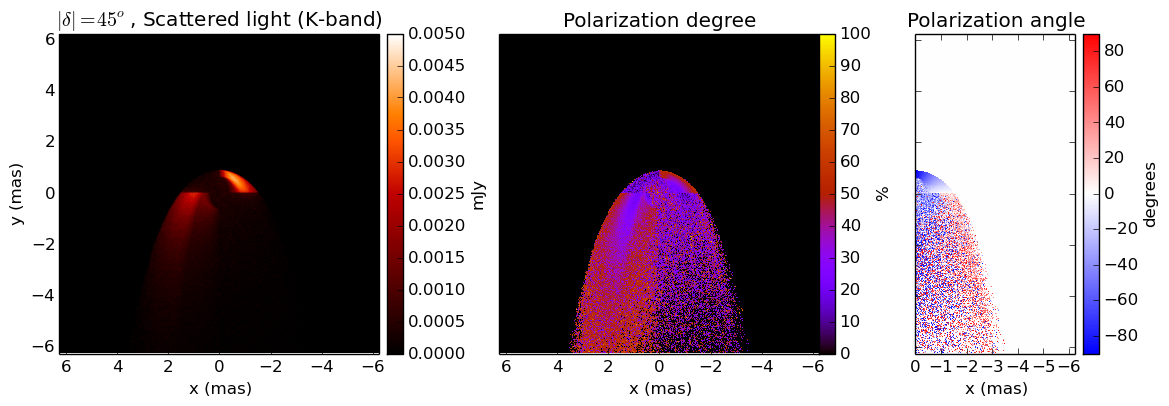}
  \includegraphics[width=\columnwidth]{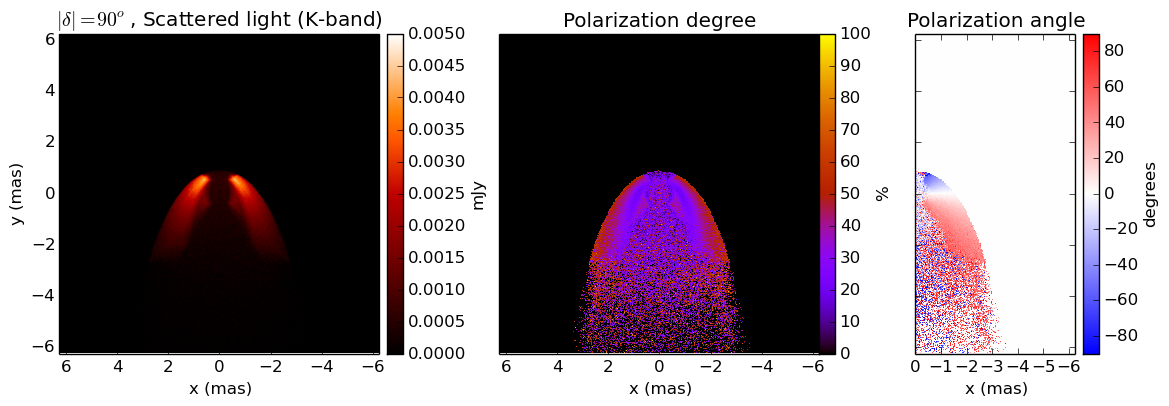}
  \caption{Emission map of scattered light in $K_{\rm{s}}$ band, the distribution of the polarization degree, and the angle in the left, middle, and right panels, respectively, for three different configurations of the star--outflow system: $\delta=0^{\circ}$, $45^{\circ}$, and $90^{\circ}$ from top to bottom.}
  \label{fig_linpol3}  
\end{figure}

For completeness, we also produce the images of scattered emission in $K_{\rm{s}}$ band, maps of linear polarization degree, and the distribution of the polarization angle for the configurations with $\delta=0^{\circ}$, $\delta=45^{\circ}$, and $\delta=90^{\circ}$ (see Fig.~\ref{fig_linpol3}). Most of the polarized flux density originates in the bow-shock layer where the bipolar stellar wind intercepts the bow shock. The change in tilt of the bipolar cavity naturally leads to the modification of the polarization angle.   

\begin{figure}[tbh]
  \centering
  \includegraphics[width=0.75\columnwidth]{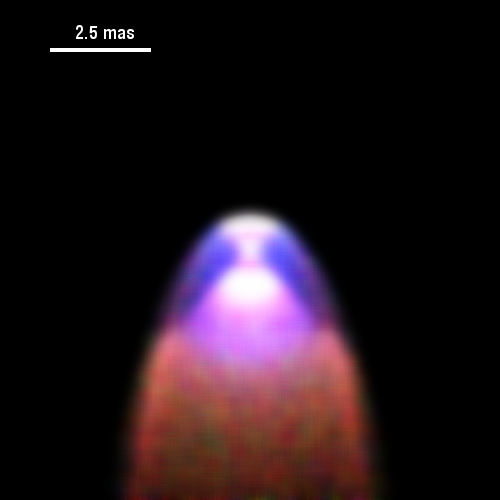}
  \caption{RGB image of the DSO source model. Blue stands for the $K_{\rm{s}}$ band, green for the $L'$ band, and red for the $M$-band emission.}
  \label{fig_rgb_dso}
\end{figure}

All in all, the main observed characteristics of the DSO, that
is, compactness, the total flux density, the polarization degree and angle, and their corresponding changes, can be reproduced by the model of a supersonic YSO. A high polarization degree in $K_{\rm{s}}$ band is due to the scattering of stellar and dust photons, and the main contribution is due to the self-scattering of dust emission. The reconstructed RGB image of the DSO model from the radiative transfer is shown in Fig.~\ref{fig_rgb_dso}, where blue stands for the $K_{\rm{s}}$ band, green for the $L'$
band, and red for the $M$-band emission.

\subsection{Effect of dichroic extinction}

A fraction of the polarized flux density of the DSO may arise from dichroic extinction. Dust grains might not be necessarily spherical, but they might be cylindrical and aligned to a certain extent by a magnetic or radiation field. A significant polarized signal may then be created by dichroic extinction: photons with electric vectors that are parallel to the grain axis experience higher extinction than those with electric vectors that are perpendicular. Therefore, a polarized emission emerges with a polarization degree $p$ that may be expressed in the following way \citep{krugel2002physics}:

\begin{equation}
  p=\frac{e^{-\tau_{\rm{min}}}-e^{-\tau_{\rm{max}}}}{e^{-\tau_{\rm{min}}}+e^{-\tau_{\rm{max}}}}\,,
\end{equation}
where $\tau_{\rm{max}}$ is the highest optical depth in the direction of the strongest attenuation and $\tau_{\rm{min}}$ is the lowest optical depth in the perpendicular direction. The difference is usually small, $\tau_{\rm{max}}-\tau_{\rm{min}}\rightarrow 0$, which then leads to
\begin{equation}
 p=\frac{1}{2}(\tau_{\rm{max}}-\tau_{\rm{min}})=\frac{1}{2} \sigma_{\rm{gr}}(Q_{\rm{ext}}^{\rm{max}}-Q_{\rm{ext}}^{\rm{min}})\int_{0}^{l} n(l) \mathrm{d}l\,,
 \label{eq_dichroic_extinction}
\end{equation}
where $\sigma_{\rm{gr}}$ is the effective cross-section of the grain, $Q_{\rm{ext}}^{\rm{max}}$ and $Q_{\rm{ext}}^{\rm{min}}$ are the highest and lowest extinction efficiencies, and the integral $N=\int_{0}^{l} n(l) \mathrm{d}l$ stands for the column density of dust along the line of sight.

Since the polarization degree $p$ in Eq.~\ref{eq_dichroic_extinction} is proportional to the column density, which tends to increase toward the pericenter as a result of the increasing surface density of the bow-shock layer (see Zaja\v{c}ek et al., 2016, for detailed calculations), the increase in polarization degree toward the pericenter (2008--2012) might be partially caused by dichroic extinction.

\section{Discussion}
\label{section:discussion}

The obtained polarization degree of the DSO in different observing years is significantly higher than the foreground polarization ($6.1\%$ in $K_\mathrm{s}$ band), which shows that it is an intrinsically polarized source. 
Emission or absorption from elongated grains that are aligned magnetically and (Mie-) scattering on dust grains in the dusty stellar envelopes are the most probable processes to produce the intrinsic polarization. In the NIR, emission and absorption processes can take place at the same time and cancel each other
out.
When scattering on the spherical dust grains occurs, part of the light that is scattered forward is unpolarized, and the rest would be scattered perpendicular to the direction of incident light, which produces linear polarization. If a central star is embedded in an isotropic envelope of spherical grains, the scattered light produces polarization with polarization vectors tangential to concentric circles about the central star that causes the overall polarization to be zero.

Polarization patterns can be more complex for more complex geometries such as YSOs that have dusty disks around the central source \citep[see, e.g.,][]{murakawa2010}, and bow-shock sources. In a system of a star that has dusty disk around it, the polarization values of different regions around the source cancel each other out when the source is viewed
face-on and no polarization would be detected, while when it is viewed edge-on, the source shows polarization \citep{Buchholz2011}.  
Scattering can also occur on elongated and non-spherical dust grains and has been modeled \citep[see, e.g.,][]{Whitney2002}, although in our radiative transfer model we assumed only spherical dust grains. All in all, polarization produced by light scattering depends on the source viewing angle and its geometry.

Since the measured polarization degree of the DSO that we obtain is high ($p>20\%$), it might be caused by non-spherical geometry, for example, by a star with a bipolar wind and a bow~shock.

In general, if the DSO is moving through the hot, X-ray emitting region of the interstellar medium with a supersonic velocity, a bow~shock can form. Collisional ionization and high densities in the shocked layer of the stellar wind can generate its observed emission lines \citep{Scoville2013}. However, no increase in the  Br$\gamma$ luminosity has been detected, which would be expected for the bow-shock dominated emission \citep{Valencia2015}. \\

The polarization angle varies during the years, that is to say, the direction of the polarization vector changes from its expected direction, which is perpendicular to the direction of motion. The modification of the polarization angle can be caused by the change in orientation of the bipolar wind of the star that occurs when the circumstellar disk is not aligned with the orbital plane. It can also be explained by a combination of the source motion and the interaction with the surrounding medium.

T~Tauri stars generally exhibit an intrinsic polarization, and one of the main origins of their polarization is scattering by dust grains in circumstellar shells \citep[see, e.g.,][]{yudin1998}. 
In polarimetric images, they show variations in polarization degree and angle \citep{Appenzeller1989}. Many T~Tauri stars have magnetospheric accretion flows since their circumstellar magnetic fields are powerful and globally ordered to maintain this type of large-scale accretion flows \citep{Symington2005}.\\

Some circumstellar disks of the low-mass stars have a bow-shock appearance like the proplyds (protoplanetary disks) of the Orion nebula cluster \citep{o'dell1994, Johnstone1998, Storzer1999}. NIR polarimetry is a tool to constrain the disk parameters of the proplyds even if the disk structure itself is not resolvable \citep{rost2008}.
Based on our data set, we cannot spatially resolve a disk or a bow-shock structure because the source appears as a point source with the current angular resolution of 8~m class telescopes in $K_\mathrm{s}$ band. It might be similar to X3 and X7, the bow-shock sources within the inner $5 ''$ at the GC \citep{muzic2010}.
Moreover, as the source is not extended, it is not possible to measure the polarization in individual regions of the source, as was done in \cite{rauch2013} for IRS 8. \\


\section{Summary and conclusions}
\label{section:summary}

We have analyzed the NIR polarized observations of a
Dusty S-cluster Object (DSO/G2) on an eccentric orbit
around Sgr A*. $K_\mathrm{s}$-band polarization data were available
for 2008, 2009, 2011, and 2012.
In all these years we clearly detected $K_\mathrm{s}$-band
continuum emission in the different channels of the Wollaston prism.
The data cover the polarization information of the DSO before its pericenter
passage (May 2014). The source does not show significant
variability in the overall $K_\mathrm{s}$-band flux density
during the observed years, and its polarization degree is mostly above
20\%, which is higher than the foreground polarization
measured on the surrounding stars. It appears that
the polarization degree is approximately constant and the polarization
angle varies as it approaches the position of Sgr A*. Based
on our significance analysis, the polarization measurements
of the DSO are significant in 2008, 2009, and 2012 and can be interpreted as source-intrinsic properties.\\

Since the total polarization degree is noticeably high, higher than $20\%$ for all epochs, the DSO structure is expected to deviate from the spherical symmetry. Moreover, the analysis of \citet{Valencia2015} showed that the source remains compact, meaning that it is not effected by tidal forces close to the pericenter of its orbit. \citet{Gillessen2012} and \citet{Eckart2013} discussed a NIR excess of $K_{\rm{s}}-L' > 3$, which implies the presence of a dense gaseous-dusty envelope. All of these basic observed parameters, the NIR excess, the compactness, and the significant polarization, may be reconciled within the model of a dust-enshrouded young star, to be precise, a pre-main-sequence star of class 1 \citep{Zajacek2014,2015wds..conf...27Z}, that forms a dense bow-shock layer by its supersonic motion upon approaching the supermassive black hole \citep{Zajacek2016}.

The obtained polarization properties of the DSO in this work can be caused by the non-spherical geometry of a bow~shock and a bipolar wind of the star. We used the 3D radiative transfer model implementing the code Hyperion \citep{2011A&A...536A..79R} to compare the observed measurements to the model of the young stellar object forming a bow~shock.

We conclude that the varying polarization angle is related to the intrinsic change of the circumstellar configuration. The change in bipolar outflow orientation may be due to the accretion disk wobbling or precession in the gravitational field of the SMBH. It can be also produced by external interaction of the DSO with the accretion flow. Although our model is simple, it can reproduce many observed properties of the DSO obtained in this work, such as the total flux density and the polarization degree.  
A more detailed analysis of the model will be provided in Zaja\v{c}ek et al., in prep.\\

\cite{shahzamanian2015b} showed that the Sgr~A* system exhibits a stable geometry and accretion process that is consistent with the preferred jet or wind directions.
The close fly-by of the DSO, or similar dusty sources (Peissker et al., in prep.), might have an effect on the stable accretion
flow onto Sgr~A* that depends on the nature of these objects. However, after the pericenter passage of the DSO, the object remained compact and its orbit Keplerian \citep{Witzel2014,Valencia2015}. Consequently, it did not lose a noticeable amount of energy and angular momentum during its closest approach to Sgr~A*, and as a result experienced weak interactions with the central black hole \citep{park2015}. However, based on hydrodynamical simulations, it may take several years for their interaction and to see a change in the activity of Sgr~A* \citep{burkert2012, schartmann2012}, either as an increase in the accretion flow rate or in the appearance of jets \citep{yuan2014}. 
Therefore, polarization and variability measurements of Sgr~A* are needed to be continued as they are
the ideal tool to probe any change in the apparently stable system as a function of the DSO fly-by. Moreover, future polarized observations of the DSO, that is, after the pericenter passage, in the NIR can help us to better constrain the source polarization and structure.\\

\begin{acknowledgements}

The authors would like to thank the anonymous referee for the helpful comments on this paper. We would like to thank G. Witzel for fruitful discussions. This work was supported in part by the Deutsche
Forschungsgemeinschaft (DFG) via the Cologne Bonn Graduate School (BCGS) and the Max Planck Society through the
International Max Planck Research School (IMPRS) for Astronomy and Astrophysics. B. Shahzamanian has been supported by
IMPRS and the BCGS. N. Sabha has been supported by BCGS. M. Zajacek and M. Parsa are
members of the IMPRS. Part of this
work was supported by fruitful discussions with members of
the Czech Science Foundation DFG collaboration (No. 13-00070J). We also received funding from
the European Union Seventh Framework Program (FP7/2007-
2013) under grant agreement n312789; Strong
gravity: Probing Strong Gravity by Black Holes Across the
Range of Masses.

\end{acknowledgements}

\vspace*{0.5cm}
\bibliographystyle{aa} 
\bibliography{ban_dso.bib} 

\begin{thebibliography}{64}
\expandafter\ifx\csname natexlab\endcsname\relax\def\natexlab#1{#1}\fi

\bibitem[{{Abarca} {et~al.}(2014){Abarca}, {S{\c a}dowski}, \&
  {Sironi}}]{Abarca2014}
{Abarca}, D., {S{\c a}dowski}, A., \& {Sironi}, L. 2014, \mnras, 440, 1125

\bibitem[{{Appenzeller} \& {Mundt}(1989)}]{Appenzeller1989}
{Appenzeller}, I. \& {Mundt}, R. 1989, \aapr, 1, 291

\bibitem[{{Bailey} {et~al.}(1984){Bailey}, {Hough}, \& {Axon}}]{Bailey1984}
{Bailey}, J., {Hough}, J.~H., \& {Axon}, D.~J. 1984, \mnras, 208, 661

\bibitem[{{Ballone} {et~al.}(2013){Ballone}, {Schartmann}, {Burkert},
  {Gillessen}, {Genzel}, {Fritz}, {Eisenhauer}, {Pfuhl}, \&
  {Ott}}]{Ballone2013}
{Ballone}, A., {Schartmann}, M., {Burkert}, A., {et~al.} 2013, \apj, 776, 13

\bibitem[{{Bate} {et~al.}(2000){Bate}, {Bonnell}, {Clarke}, {Lubow}, {Ogilvie},
  {Pringle}, \& {Tout}}]{2000MNRAS.317..773B}
{Bate}, M.~R., {Bonnell}, I.~A., {Clarke}, C.~J., {et~al.} 2000, \mnras, 317,
  773

\bibitem[{{Brandner} {et~al.}(2002){Brandner}, {Rousset}, {Lenzen}, {Hubin},
  {Lacombe}, {Hofmann}, {Moorwood}, {Lagrange}, {Gendron}, {Hartung}, {Puget},
  {Ageorges}, {Biereichel}, {Bouy}, {Charton}, {Dumont}, {Fusco}, {Jung},
  {Lehnert}, {Lizon}, {Monnet}, {Mouillet}, {Moutou}, {Rabaud}, {R{\"o}hrle},
  {Skole}, {Spyromilio}, {Storz}, {Tacconi-Garman}, \& {Zins}}]{Brandner2002}
{Brandner}, W., {Rousset}, G., {Lenzen}, R., {et~al.} 2002, The Messenger, 107,
  1

\bibitem[{{Buchholz} {et~al.}(2013){Buchholz}, {Witzel}, {Sch{\"o}del}, \&
  {Eckart}}]{Buchholz2013}
{Buchholz}, R.~M., {Witzel}, G., {Sch{\"o}del}, R., \& {Eckart}, A. 2013, \aap,
  557, A82

\bibitem[{{Buchholz} {et~al.}(2011){Buchholz}, {Witzel}, {Sch{\"o}del},
  {Eckart}, {Bremer}, \& {Mu{\v z}i{\'c}}}]{Buchholz2011}
{Buchholz}, R.~M., {Witzel}, G., {Sch{\"o}del}, R., {et~al.} 2011, \aap, 534,
  A117

\bibitem[{{Burkert} {et~al.}(2012){Burkert}, {Schartmann}, {Alig}, {Gillessen},
  {Genzel}, {Fritz}, \& {Eisenhauer}}]{burkert2012}
{Burkert}, A., {Schartmann}, M., {Alig}, C., {et~al.} 2012, \apj, 750, 58

\bibitem[{{Devillard}(1999)}]{Devillard1999}
{Devillard}, N. 1999, in Astronomical Society of the Pacific Conference Series,
  Vol. 172, Astronomical Data Analysis Software and Systems VIII, ed. D.~M.
  {Mehringer}, R.~L. {Plante}, \& D.~A. {Roberts}, 333

\bibitem[{{Diolaiti} {et~al.}(2000){Diolaiti}, {Bendinelli}, {Bonaccini},
  {Close}, {Currie}, \& {Parmeggiani}}]{Diolaiti2000}
{Diolaiti}, E., {Bendinelli}, O., {Bonaccini}, D., {et~al.} 2000, \aaps, 147,
  335

\bibitem[{{Eckart} {et~al.}(2012){Eckart}, {Britzen}, {Horrobin},
  {Zamaninasab}, {Muzic}, {Sabha}, {Shahzamanian}, {Yazici}, {Moser}, {Zuther},
  {Garcia-Marin}, {Valencia-S.}, {Bursa}, {Karssen}, {Karas}, {Jalali},
  {Vitale}, {Bremer}, {Fischer}, {Smajic}, {Rauch}, {Kunneriath}, {Moultaka},
  {Straubmeier}, {Rashed}, {Iserlohe}, {Busch}, {Markakis}, {Borkar}, \&
  {Zensus}}]{eckart2012a}
{Eckart}, A., {Britzen}, S., {Horrobin}, M., {et~al.} 2012, in Proceedings of
  Nuclei of Seyfert galaxies and QSOs - Central engine \& conditions of star
  formation (Seyfert 2012). 6-8 November, 2012. Max-Planck-Insitut f{\"u}r
  Radioastronomie (MPIfR), Bonn, Germany., 4

\bibitem[{{Eckart} {et~al.}(2013){Eckart}, {Mu{\v z}i{\'c}}, {Yazici}, {Sabha},
  {Shahzamanian}, {Witzel}, {Moser}, {Garcia-Marin}, {Valencia-S.}, {Jalali},
  {Bremer}, {Straubmeier}, {Rauch}, {Buchholz}, {Kunneriath}, \&
  {Moultaka}}]{Eckart2013}
{Eckart}, A., {Mu{\v z}i{\'c}}, K., {Yazici}, S., {et~al.} 2013, \aap, 551, A18

\bibitem[{{Ghez} {et~al.}(2014){Ghez}, {Witzel}, {Sitarski}, {Meyer}, {Yelda},
  {Boehle}, {Becklin}, {Campbell}, {Canalizo}, {Do}, {Lu}, {Matthews},
  {Morris}, \& {Stockton}}]{Ghez2014}
{Ghez}, A.~M., {Witzel}, G., {Sitarski}, B., {et~al.} 2014, The Astronomer's
  Telegram, 6110, 1

\bibitem[{{Gillessen} {et~al.}(2013){Gillessen}, {Genzel}, {Fritz},
  {Eisenhauer}, {Pfuhl}, {Ott}, {Cuadra}, {Schartmann}, \&
  {Burkert}}]{Gillessen2013a}
{Gillessen}, S., {Genzel}, R., {Fritz}, T.~K., {et~al.} 2013, \apj, 763, 78

\bibitem[{{Gillessen} {et~al.}(2012){Gillessen}, {Genzel}, {Fritz}, {Quataert},
  {Alig}, {Burkert}, {Cuadra}, {Eisenhauer}, {Pfuhl}, {Dodds-Eden}, {Gammie},
  \& {Ott}}]{Gillessen2012}
{Gillessen}, S., {Genzel}, R., {Fritz}, T.~K., {et~al.} 2012, \nat, 481, 51

\bibitem[{{Jalali} {et~al.}(2014){Jalali}, {Pelupessy}, {Eckart}, {Portegies
  Zwart}, {Sabha}, {Borkar}, {Moultaka}, {Mu{\v z}i{\'c}}, \&
  {Moser}}]{Jalali2014}
{Jalali}, B., {Pelupessy}, F.~I., {Eckart}, A., {et~al.} 2014, \mnras, 444,
  1205

\bibitem[{{Johnstone} {et~al.}(1998){Johnstone}, {Hollenbach}, \&
  {Bally}}]{Johnstone1998}
{Johnstone}, D., {Hollenbach}, D., \& {Bally}, J. 1998, \apj, 499, 758

\bibitem[{{Kim} {et~al.}(1994){Kim}, {Martin}, \&
  {Hendry}}]{1994ApJ...422..164K}
{Kim}, S.-H., {Martin}, P.~G., \& {Hendry}, P.~D. 1994, \apj, 422, 164

\bibitem[{Krugel(2002)}]{krugel2002physics}
Krugel, E. 2002, The Physics of Interstellar Dust, Series in Astronomy and
  Astrophysics (Taylor \& Francis)

\bibitem[{{Lenzen} {et~al.}(2003){Lenzen}, {Hartung}, {Brandner}, {Finger},
  {Hubin}, {Lacombe}, {Lagrange}, {Lehnert}, {Moorwood}, \&
  {Mouillet}}]{Lenzen2003}
{Lenzen}, R., {Hartung}, M., {Brandner}, W., {et~al.} 2003, in Society of
  Photo-Optical Instrumentation Engineers (SPIE) Conference Series, Vol. 4841,
  Instrument Design and Performance for Optical/Infrared Ground-based
  Telescopes, ed. M.~{Iye} \& A.~F.~M. {Moorwood}, 944--952

\bibitem[{{Mapelli} \& {Trani}(2016)}]{mapelli2016}
{Mapelli}, M. \& {Trani}, A.~A. 2016, \aap, 585, A161

\bibitem[{{Meyer} {et~al.}(2014{\natexlab{a}}){Meyer}, {Ghez}, {Witzel}, {Do},
  {Phifer}, {Sitarski}, {Morris}, {Boehle}, {Yelda}, {Lu}, \&
  {Becklin}}]{Meyer2014a}
{Meyer}, L., {Ghez}, A.~M., {Witzel}, G., {et~al.} 2014{\natexlab{a}}, in IAU
  Symposium, Vol. 303, IAU Symposium, ed. L.~O. {Sjouwerman}, C.~C. {Lang}, \&
  J.~{Ott}, 264--268

\bibitem[{{Meyer} {et~al.}(2014{\natexlab{b}}){Meyer}, {Witzel}, {Longstaff},
  \& {Ghez}}]{meyer2014}
{Meyer}, L., {Witzel}, G., {Longstaff}, F.~A., \& {Ghez}, A.~M.
  2014{\natexlab{b}}, \apj, 791, 24

\bibitem[{{Monnier} \& {Millan-Gabet}(2002)}]{2002ApJ...579..694M}
{Monnier}, J.~D. \& {Millan-Gabet}, R. 2002, \apj, 579, 694

\bibitem[{{Murakawa}(2010)}]{murakawa2010}
{Murakawa}, K. 2010, \aap, 518, A63

\bibitem[{{Murray-Clay} \& {Loeb}(2012)}]{Murray-Clay2012}
{Murray-Clay}, R.~A. \& {Loeb}, A. 2012, Nature Communications, 3, 1049

\bibitem[{{Mu{\v z}i{\'c}} {et~al.}(2010){Mu{\v z}i{\'c}}, {Eckart},
  {Sch{\"o}del}, {Buchholz}, {Zamaninasab}, \& {Witzel}}]{muzic2010}
{Mu{\v z}i{\'c}}, K., {Eckart}, A., {Sch{\"o}del}, R., {et~al.} 2010, \aap,
  521, A13

\bibitem[{{O'dell} \& {Wen}(1994)}]{o'dell1994}
{O'dell}, C.~R. \& {Wen}, Z. 1994, \apj, 436, 194

\bibitem[{{Park} {et~al.}(2015){Park}, {Trippe}, {Krichbaum}, {Kim}, {Kino},
  {Bertarini}, {Bremer}, \& {de Vicente}}]{park2015}
{Park}, J.-H., {Trippe}, S., {Krichbaum}, T.~P., {et~al.} 2015, \aap, 576, L16

\bibitem[{{Pfuhl} {et~al.}(2015){Pfuhl}, {Gillessen}, {Eisenhauer}, {Genzel},
  {Plewa}, {Ott}, {Ballone}, {Schartmann}, {Burkert}, {Fritz}, {Sari},
  {Steinberg}, \& {Madigan}}]{Pfuhl2015}
{Pfuhl}, O., {Gillessen}, S., {Eisenhauer}, F., {et~al.} 2015, \apj, 798, 111

\bibitem[{{Phifer} {et~al.}(2013){Phifer}, {Do}, {Meyer}, {Ghez}, {Witzel},
  {Yelda}, {Boehle}, {Lu}, {Morris}, {Becklin}, \& {Matthews}}]{Phifer2013}
{Phifer}, K., {Do}, T., {Meyer}, L., {et~al.} 2013, \apjl, 773, L13

\bibitem[{{Rauch} {et~al.}(2013){Rauch}, {Mu{\v z}i{\'c}}, {Eckart},
  {Buchholz}, {Garc{\'{\i}}a-Mar{\'{\i}}n}, {Sabha}, {Straubmeier},
  {Valencia-S.}, \& {Yazici}}]{rauch2013}
{Rauch}, C., {Mu{\v z}i{\'c}}, K., {Eckart}, A., {et~al.} 2013, \aap, 551, A35

\bibitem[{{Robitaille}(2011)}]{2011A&A...536A..79R}
{Robitaille}, T.~P. 2011, \aap, 536, A79

\bibitem[{{Rost} {et~al.}(2008){Rost}, {Eckart}, \& {Ott}}]{rost2008}
{Rost}, S., {Eckart}, A., \& {Ott}, T. 2008, \aap, 485, 107

\bibitem[{{Rousset} {et~al.}(2003){Rousset}, {Lacombe}, {Puget}, {Hubin},
  {Gendron}, {Fusco}, {Arsenault}, {Charton}, {Feautrier}, {Gigan}, {Kern},
  {Lagrange}, {Madec}, {Mouillet}, {Rabaud}, {Rabou}, {Stadler}, \&
  {Zins}}]{Rousset2003}
{Rousset}, G., {Lacombe}, F., {Puget}, P., {et~al.} 2003, in Society of
  Photo-Optical Instrumentation Engineers (SPIE) Conference Series, Vol. 4839,
  Adaptive Optical System Technologies II, ed. P.~L. {Wizinowich} \&
  D.~{Bonaccini}, 140--149

\bibitem[{{Sabha} {et~al.}(2012){Sabha}, {Eckart}, {Merritt}, {Zamaninasab},
  {Witzel}, {Garc{\'{\i}}a-Mar{\'{\i}}n}, {Jalali}, {Valencia-S.}, {Yazici},
  {Buchholz}, {Shahzamanian}, {Rauch}, {Horrobin}, \&
  {Straubmeier}}]{sabha2012}
{Sabha}, N., {Eckart}, A., {Merritt}, D., {et~al.} 2012, \aap, 545, A70

\bibitem[{{Sanchez-Bermudez} {et~al.}(2016){Sanchez-Bermudez}, {Hummel},
  {Tuthill}, {Alberdi}, {Sch{\"o}del}, {Lacour}, \& {Stanke}}]{Sanchez2016}
{Sanchez-Bermudez}, J., {Hummel}, C.~A., {Tuthill}, P., {et~al.} 2016, \aap,
  588, A117

\bibitem[{{S{\c a}dowski} {et~al.}(2013){S{\c a}dowski}, {Sironi}, {Abarca},
  {Guo}, {{\"O}zel}, \& {Narayan}}]{Sadowski2013}
{S{\c a}dowski}, A., {Sironi}, L., {Abarca}, D., {et~al.} 2013, \mnras, 432,
  478

\bibitem[{{Schartmann} {et~al.}(2012){Schartmann}, {Burkert}, {Alig},
  {Gillessen}, {Genzel}, {Eisenhauer}, \& {Fritz}}]{schartmann2012}
{Schartmann}, M., {Burkert}, A., {Alig}, C., {et~al.} 2012, \apj, 755, 155

\bibitem[{{Sch{\"o}del} {et~al.}(2010){Sch{\"o}del}, {Najarro}, {Muzic}, \&
  {Eckart}}]{schoedel2010}
{Sch{\"o}del}, R., {Najarro}, F., {Muzic}, K., \& {Eckart}, A. 2010, \aap, 511,
  A18

\bibitem[{{Scoville} \& {Burkert}(2013)}]{Scoville2013}
{Scoville}, N. \& {Burkert}, A. 2013, \apj, 768, 108

\bibitem[{{Serkowski}(1958)}]{serkowski1958}
{Serkowski}, K. 1958, \actaa, 8, 135

\bibitem[{{Shahzamanian} {et~al.}(2015){Shahzamanian}, {Eckart}, {Valencia-S.},
  {Witzel}, {Zamaninasab}, {Sabha}, {Garc{\'{\i}}a-Mar{\'{\i}}n}, {Karas},
  {Karssen}, {Borkar}, {Dov{\v c}iak}, {Kunneriath}, {Bursa}, {Buchholz},
  {Moultaka}, \& {Straubmeier}}]{shahzamanian2015b}
{Shahzamanian}, B., {Eckart}, A., {Valencia-S.}, M., {et~al.} 2015, \aap, 576,
  A20

\bibitem[{{Shcherbakov}(2014)}]{Shcherbakov2014}
{Shcherbakov}, R.~V. 2014, \apj, 783, 31

\bibitem[{{Shcherbakov} \& {Baganoff}(2010)}]{shcherbakov2010}
{Shcherbakov}, R.~V. \& {Baganoff}, F.~K. 2010, \apj, 716, 504

\bibitem[{{Simmons} \& {Stewart}(1985)}]{simmons1985}
{Simmons}, J.~F.~L. \& {Stewart}, B.~G. 1985, \aap, 142, 100

\bibitem[{{St{\"o}rzer} \& {Hollenbach}(1999)}]{Storzer1999}
{St{\"o}rzer}, H. \& {Hollenbach}, D. 1999, \apj, 515, 669

\bibitem[{{Symington} {et~al.}(2005){Symington}, {Harries}, \&
  {Kurosawa}}]{Symington2005}
{Symington}, N.~H., {Harries}, T.~J., \& {Kurosawa}, R. 2005, \mnras, 356, 1489

\bibitem[{{Valencia-S.} {et~al.}(2015){Valencia-S.}, {Eckart}, {Zaja{\v c}ek},
  {Peissker}, {Parsa}, {Grosso}, {Mossoux}, {Porquet}, {Jalali}, {Karas},
  {Yazici}, {Shahzamanian}, {Sabha}, {Saalfeld}, {Smajic}, {Grellmann},
  {Moser}, {Horrobin}, {Borkar}, {Garc{\'{\i}}a-Mar{\'{\i}}n}, {Dov{\v c}iak},
  {Kunneriath}, {Karssen}, {Bursa}, {Straubmeier}, \&
  {Bushouse}}]{Valencia2015}
{Valencia-S.}, M., {Eckart}, A., {Zaja{\v c}ek}, M., {et~al.} 2015, \apj, 800,
  125

\bibitem[{{Vinokur}(1965)}]{vinokur1965}
{Vinokur}, M. 1965, Annales d'Astrophysique, 28, 412

\bibitem[{{Whitney} \& {Wolff}(2002)}]{Whitney2002}
{Whitney}, B.~A. \& {Wolff}, M.~J. 2002, \apj, 574, 205

\bibitem[{{Wilkin}(1996)}]{wilkin1996}
{Wilkin}, F.~P. 1996, \apjl, 459, L31

\bibitem[{{Witzel} {et~al.}(2012){Witzel}, {Eckart}, {Bremer}, {Zamaninasab},
  {Shahzamanian}, {Valencia-S.}, {Sch{\"o}del}, {Karas}, {Lenzen}, {Marchili},
  {Sabha}, {Garcia-Marin}, {Buchholz}, {Kunneriath}, \&
  {Straubmeier}}]{witzel2012}
{Witzel}, G., {Eckart}, A., {Bremer}, M., {et~al.} 2012, \apjs, 203, 18

\bibitem[{{Witzel} {et~al.}(2011){Witzel}, {Eckart}, {Buchholz}, {Zamaninasab},
  {Lenzen}, {Sch{\"o}del}, {Araujo}, {Sabha}, {Bremer}, {Karas}, {Straubmeier},
  \& {Muzic}}]{Witzel2011}
{Witzel}, G., {Eckart}, A., {Buchholz}, R.~M., {et~al.} 2011, \aap, 525, A130

\bibitem[{{Witzel} {et~al.}(2014){Witzel}, {Ghez}, {Morris}, {Sitarski},
  {Boehle}, {Naoz}, {Campbell}, {Becklin}, {Canalizo}, {Chappell}, {Do}, {Lu},
  {Matthews}, {Meyer}, {Stockton}, {Wizinowich}, \& {Yelda}}]{Witzel2014}
{Witzel}, G., {Ghez}, A.~M., {Morris}, M.~R., {et~al.} 2014, \apjl, 796, L8

\bibitem[{{Yuan} \& {Narayan}(2014)}]{yuan2014}
{Yuan}, F. \& {Narayan}, R. 2014, \araa, 52, 529

\bibitem[{{Yudin} \& {Evans}(1998)}]{yudin1998}
{Yudin}, R.~V. \& {Evans}, A. 1998, \aaps, 131, 401

\bibitem[{{Yusef-Zadeh} {et~al.}(2015{\natexlab{a}}){Yusef-Zadeh}, {Bushouse},
  {Sch{\"o}del}, {Wardle}, {Cotton}, {Roberts}, {Nogueras-Lara}, \&
  {Gallego-Cano}}]{yusef2015}
{Yusef-Zadeh}, F., {Bushouse}, H., {Sch{\"o}del}, R., {et~al.}
  2015{\natexlab{a}}, \apj, 809, 10

\bibitem[{{Yusef-Zadeh} {et~al.}(2015{\natexlab{b}}){Yusef-Zadeh}, {Roberts},
  {Wardle}, {Cotton}, {Sch{\"o}del}, \& {Royster}}]{2015ApJ...801L..26Y}
{Yusef-Zadeh}, F., {Roberts}, D.~A., {Wardle}, M., {et~al.} 2015{\natexlab{b}},
  \apjl, 801, L26

\bibitem[{{Zajacek} {et~al.}(2015){Zajacek}, {Eckart}, {Peissker}, {Karssen},
  \& {Karas}}]{2015wds..conf...27Z}
{Zajacek}, M., {Eckart}, A., {Peissker}, F., {Karssen}, G.~D., \& {Karas}, V.
  2015, in Proceedings of the 24th Annual Conference of Doctoral Students - WDS
  2015 - Physics (eds. J. Safrankova and J. Pavlu), Prague, Matfyzpress, pp.
  27-35, 2015; ISBN 978-80-7378-311-2, 27--35

\bibitem[{{Zaja{\v c}ek} {et~al.}(2016){Zaja{\v c}ek}, {Eckart}, {Karas},
  {Kunneriath}, {Shahzamanian}, {Sabha}, {Mu{\v z}i{\'c}}, \&
  {Valencia-S.}}]{Zajacek2016}
{Zaja{\v c}ek}, M., {Eckart}, A., {Karas}, V., {et~al.} 2016, \mnras, 455, 1257

\bibitem[{{Zaja{\v c}ek} {et~al.}(2014){Zaja{\v c}ek}, {Karas}, \&
  {Eckart}}]{Zajacek2014}
{Zaja{\v c}ek}, M., {Karas}, V., \& {Eckart}, A. 2014, \aap, 565, A17

\bibitem[{{Zhang} \& {Zheng}(1997)}]{1997ApJ...474..719Z}
{Zhang}, Q. \& {Zheng}, X. 1997, \apj, 474, 719

\end{thebibliography}
 \begin{appendix}
 \section{}
We present the high-pass-filtered (smooth-subtracted) images of the central 0.65''$\times$0.65'' for 2008 and 2012 in Fig.~A.1. The DSO can be identified as a source component in the images.

    \begin{figure*}
     \begin{center}

        \subfigure{%
           \includegraphics[width=0.45\textwidth]{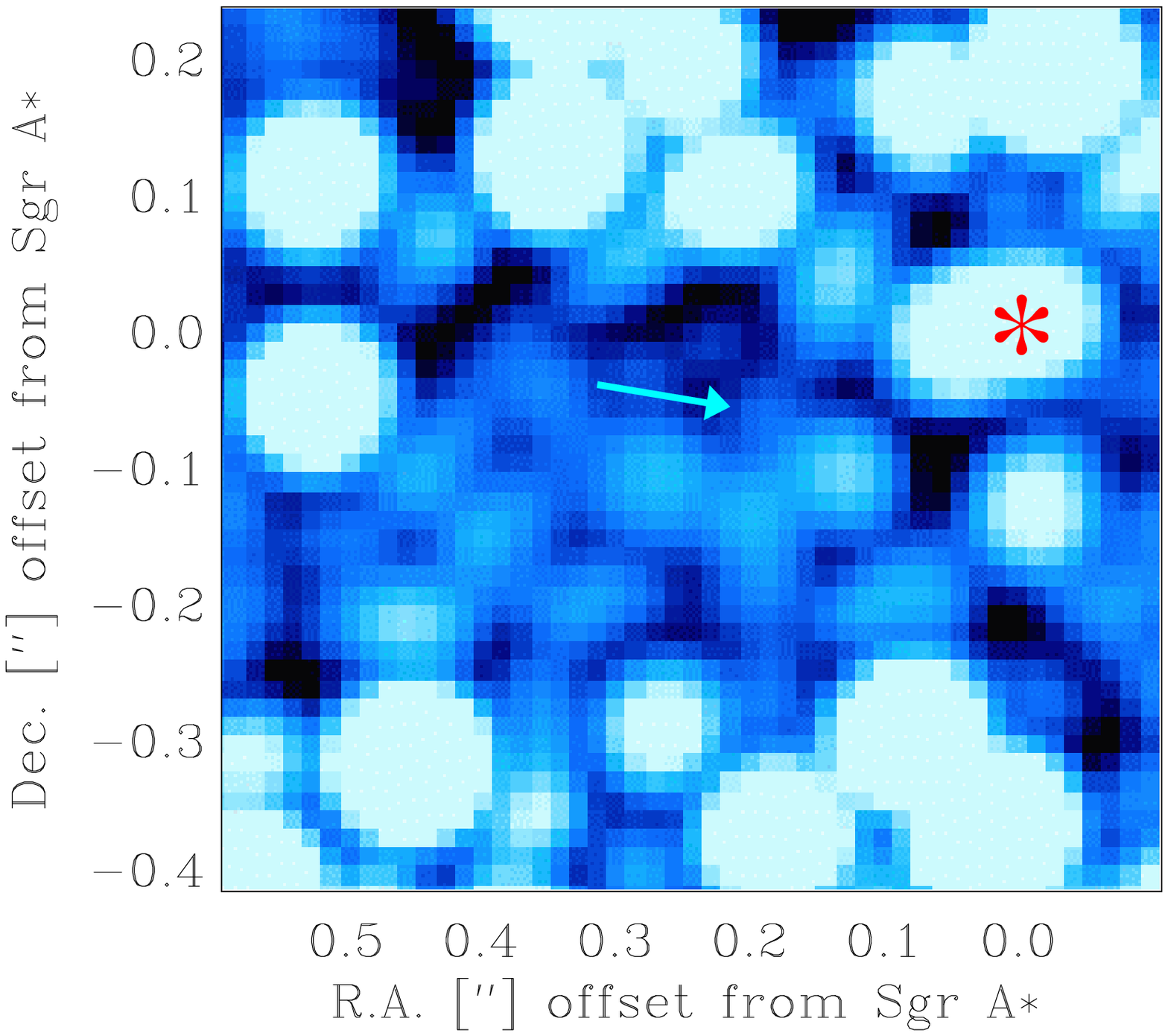}
        }
        \subfigure{%
            \includegraphics[width=0.45\textwidth]{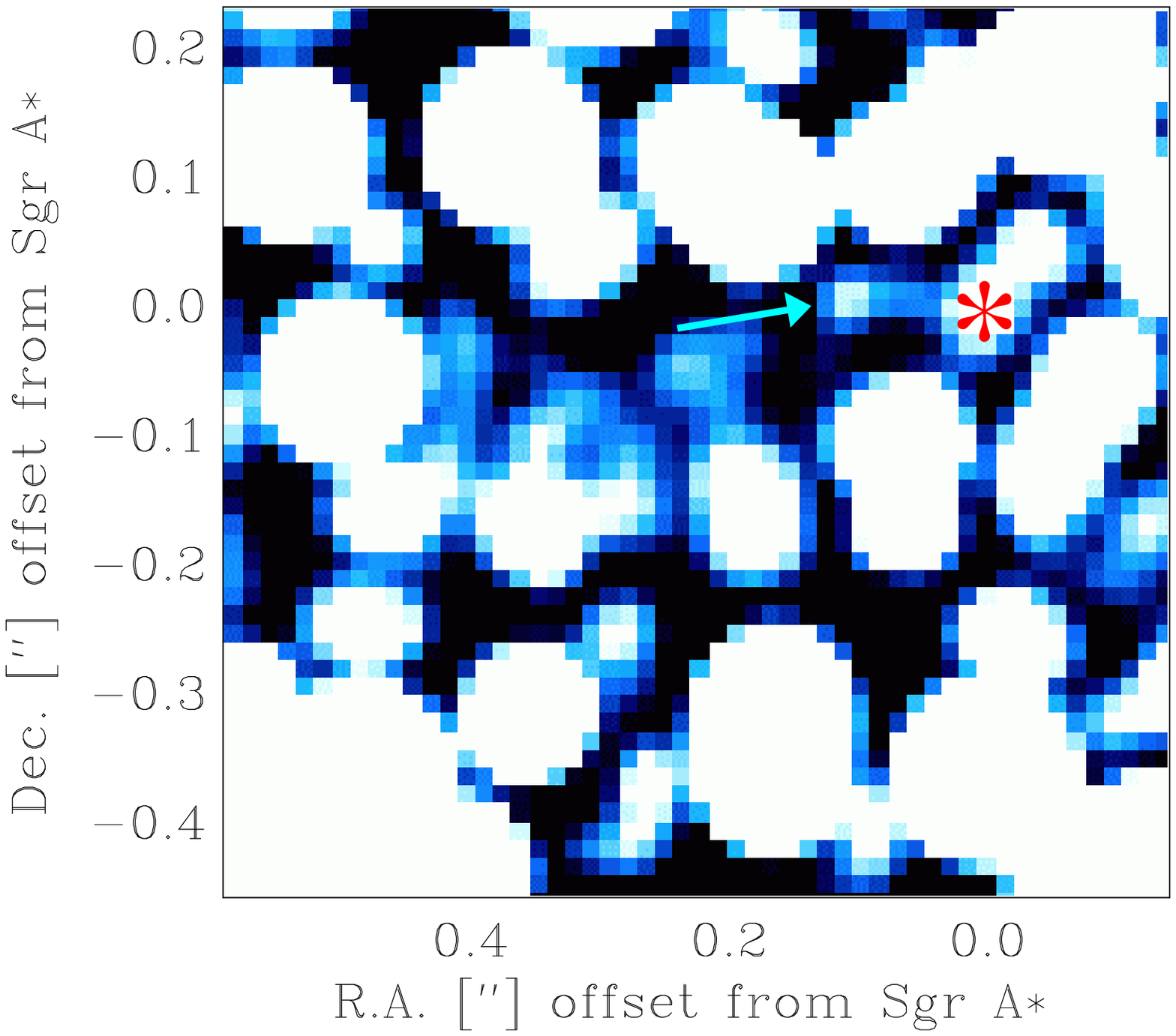}
        }\\

    \end{center}
    \caption[$K_\mathrm{s}$-band deconvolved median images of the central arcsecond at the GC in polarimetry mode]{Color maps of high-pass-filtered (smooth-subtracted) adaptive optics images
of the central 0.65''$\times$0.65''.
The arrow points at the contour line excursions that
are due to the flux density contribution of the DSO.
The red asterisk indicates the position of Sgr~A*.
Left is 90$^o$ position angle for 2008 and right is 90$^o$ position angle for 2012.}

\label{fig:dso-highpass}
\end{figure*}


\end{appendix}
\end{document}